\documentclass[11pt,oneside]{article}
\usepackage{setspace,graphicx,amssymb,amsmath,latexsym,amsfonts,amscd,amsthm,multirow,ctable,mathdots,caption,array}

\usepackage{fancyhdr,tabularx,cite,mathrsfs,color}
\usepackage[headings]{fullpage}
\usepackage{stmaryrd}
\usepackage{rotating}
\usepackage{authblk}
\usepackage{hyperref}
\usepackage{accents}

\usepackage{geometry}

\newcommand{\ubar}[1]{\underaccent{\bar}{#1}}
\newcommand{\bea}{\begin{eqnarray}} 
\newcommand{\eea}{\end{eqnarray}} 
\newcommand{\bee}{\begin{eqnarray*}} 
\newcommand{\eee}{\end{eqnarray*}} 
\newcommand{\al}{\begin{align*}} 
\newcommand{\eal}{\end{align*}} 
\newcommand{\be}{\begin{equation}} 
\newcommand{\ee}{\end{equation}} 
\newcommand{\eq}[1]{(\ref{#1})} 
\newcommand{\bem}{\begin{pmatrix}} 
\newcommand{\eem}{\end{pmatrix}} 

\def\a{\alpha} 
\def\b{\beta} 
\def\c{\gamma}

\def\f{\phi}

\def\l{\lambda} 
\def\m{\mu}

\def\p{\pi}    
\def\pa{\partial}        

\def\r{\rho}                  
            
\def\t{\tau} 
\def\th{\theta}

\def\z{\zeta}

\def\L{\Lambda}

\newcolumntype{R}{ >{$}r <{$}}
\newcolumntype{C}{ >{$}c <{$}}
\newcolumntype{L}{ >{$}l <{$}}
\newcolumntype{F}{>{\centering\arraybackslash}m{1.5cm}}

\newcommand{\comment}[1]{}
\newcommand{\grpZ}{{\mathbb Z}_}

\newcommand{\RR}{{\mathbb R}}%Reals
\newcommand{\CC}{{\mathbb C}}%Complex
\newcommand{\PP}{{\mathbb P}}%Projective
\newcommand{\ZZ}{{\mathbb Z}}%Integers
\newcommand{\G}{\Gamma}	%Gamma
\newcommand{\Co}{\textsl{Co}}	%Conway group
\newcommand{\xmod}{{\rm \;mod\;}}

\theoremstyle{definition}

\theoremstyle{remark}

\numberwithin{equation}{section}
\makeatletter
\newsavebox\myboxA
\newsavebox\myboxB
\newlength\mylenA

\newcommand*\xunderline[2][0.75]{%
    \sbox{\myboxA}{$\m@th#2$}%
    \setbox\myboxB\null% Phantom box
    \ht\myboxB=\ht\myboxA%
    \dp\myboxB=\dp\myboxA%
    \wd\myboxB=#1\wd\myboxA% Scale phantom
    \sbox\myboxB{$\m@th\underline{\copy\myboxB}$}%  Overlined phantom
    \setlength\mylenA{\the\wd\myboxA}%   calc width diff
    \addtolength\mylenA{-\the\wd\myboxB}%
    \ifdim\wd\myboxB<\wd\myboxA%
       \rlap{\hskip 0.5\mylenA\usebox\myboxB}{\usebox\myboxA}%
    \else
        \hskip -0.5\mylenA\rlap{\usebox\myboxA}{\hskip 0.5\mylenA\usebox\myboxB}%
    \fi}
\makeatother

\makeatletter
\newsavebox\myboxC
\newsavebox\myboxD
\newlength\mylenC

\newcommand*\xoverline[2][0.75]{%
    \sbox{\myboxC}{$\m@th#2$}%
    \setbox\myboxD\null% Phantom box
    \ht\myboxD=\ht\myboxC%
    \dp\myboxD=\dp\myboxC%
    \wd\myboxD=#1\wd\myboxC% Scale phantom
    \sbox\myboxD{$\m@th\overline{\copy\myboxD}$}%  Overlined phantom
    \setlength\mylenC{\the\wd\myboxC}%   calc width diff
    \addtolength\mylenC{-\the\wd\myboxD}%
    \ifdim\wd\myboxD<\wd\myboxC%
       \rlap{\hskip 0.5\mylenC\usebox\myboxD}{\usebox\myboxC}%
    \else
        \hskip -0.5\mylenC\rlap{\usebox\myboxC}{\hskip 0.5\mylenC\usebox\myboxD}%
    \fi}
\makeatother

\makeatletter
\renewcommand*{\@fnsymbol}[1]{\ifcase#1\or$\heartsuit$\or$\diamondsuit$\or$\clubsuit$\or$\spadesuit$\or$\S$\or\else\@arabic{#1}\fi}
\makeatother

\begin{document}

\setstretch{1.4}

\title{
\vspace{-35pt}
    \textsc{\huge{Landau-Ginzburg Orbifolds and Symmetries of K3 CFTs
    }  }
}

\author[1,2]{\small{Miranda C. N. Cheng}\thanks{mcheng@uva.nl (On leave from CNRS, France.)}}
\author[2]{\small{Francesca Ferrari}\thanks{f.ferrari@uva.nl}}
\author[3]{\small{Sarah M. Harrison}\thanks{sarharr@physics.harvard.edu}}
\author[4]{\small{Natalie M. Paquette}\thanks{npaquett@stanford.edu}}

\date{}

\affil[1]{Korteweg-de Vries Institute for Mathematics, Amsterdam, the Netherlands}
\affil[2]{Institute of Physics, University of Amsterdam, Amsterdam, the Netherlands}
\affil[3]{Center for the Fundamental Laws of Nature, 
Harvard University, Cambridge, MA 02138, USA}
\affil[4]{Stanford Institute for Theoretical Physics, Department of Physics\\
and Theory Group, SLAC \\
Stanford University, Stanford, CA 94305, USA\vspace{-15pt}}

   \maketitle
\abstract{
Recent developments in the study of the moonshine phenomenon, including umbral and Conway moonshine, suggest that it may play an important role in encoding the action of finite symmetry groups on the BPS spectrum of $K3$ string theory.
To test and clarify these proposed $K3$-moonshine connections, we study Landau-Ginzburg orbifolds that flow to conformal field theories in the moduli space of $K3$ sigma models.  We compute $K3$ elliptic genera twined by discrete symmetries that are manifest in the UV description, though often inaccessible in the IR.
We obtain various twining functions coinciding with moonshine predictions that have not been observed in physical theories before. 
These include twining functions arising from Mathieu moonshine, other cases of umbral moonshine, and Conway moonshine. 
For instance, all functions arising from $M_{11} \subset 2.M_{12}$  moonshine appear as explicit twining genera in the LG models,  which moreover admit a uniform description in terms of its natural 12-dimensional representation. 
Our results provide strong evidence for the relevance of umbral moonshine for $K3$ symmetries, as well as new hints for its eventual explanation. 
}

\newpage
  \tableofcontents

  \newpage
  \section{Introduction}

Five years after the observation relating the elliptic genus of $K3$ and the sporadic group $M_{24}$ \cite{Eguchi2010}, the mystery of $M_{24}$ moonshine remains. 
In the meantime, great progress has been made in the understanding of both the nature of this type of moonshine and the symmetries of $K3$ sigma models and $K3$ string theory in general. 
See \cite{Taormina:2011rr,Gaberdiel:2012gf,Gaberdiel:2012um,Gaberdiel:2010ca,Cheng:2013kpa,Gaberdiel:2013psa,Taormina:2013mda,Harvey:2013mda,Taormina:2013jza,Persson:2013xpa,Gaberdiel:2013nya,Cheng:2014owa,Harvey:2014cva,Duncan:2014vfa,Benjamin:2014kna,Wrase:2014fja,Paquette:2014rma,Cheng:2015fha,UM,umbraltwo,MR3357517,Creutzig2012,MR3271177,DuncanMackCrane,KHP}.

In the former category, it was realized that  $M_{24}$ moonshine is but one out of  23 cases of the so-called umbral moonshine  \cite{UM,umbraltwo}. 
The main data to describe umbral moonshine consist of the 23 Niemeier lattices $N(X)$, each of them uniquely determined by its root system $X$, which is one of the 23 unions of $ADE$ root systems of the same Coxeter number with total rank 24. Recall that (up to isomorphism) there are exactly 24 rank 24, positive definite, self-dual, and even lattices. These include the 23 Niemeier lattices as well as the Leech lattice which does not contain any roots (lattice vectors with norm square 2). To each of the 23 Niemeier lattice $N(X)$ we associate a finite group, the ``umbral group" $G(X)$, defined as the automorphism of $N(X)$ modded out by the Weyl group of the root system $X$. 
At the same time, using recent results on mock modular forms, we uniquely associate a mock Jacobi form $\psi^X_g$ with special properties to every conjugacy class $[g]$ of the group $G(X)$. Given the group $G(X)$ and
the functions $\psi^X_g$, the umbral moonshine conjecture states that there exists a naturally defined infinite-dimensional module for $G(X)$ such that the Fourier expansion of $\psi^X_g$ is nothing other than the (graded) $g$-character of this module. The existence of such a module has been proven in \cite{Gannon:2012ck,Duncan:2015rga}. 
The case of umbral moonshine with the simplest root system $X=24A_1$ reproduces $M_{24}$ moonshine (note that $G({24A_1}) \cong  M_{24}$) \cite{Eguchi2010,ChengM24K3,Gaberdiel2010,Gaberdiel:2010ca,Eguchi2010a,Cheng:2011ay,Cheng:2012uy}. 

A natural question is: what is the physical context of umbral moonshine? In particular, 
given the (at the very least historical) relation between $M_{24}$ moonshine and  $K3$ CFTs, one might wonder whether the other 22 cases of moonshine also share a relation to   $K3$ CFTs. 
In part inspired by the important role of Niemeier lattices in describing the geometric symmetries of $K3$ surfaces \cite{Nikulin:2011}, such a relation was proposed in \cite{UMk3}. In particular, for each of the 23 $X$ and $g\in G(X)$, a (weight 0, weak) Jacobi form $\f^X_g$, constructed in a simple and uniform way from the mock Jacobi form $\psi^X_g$, was proposed to play the role of the twined $K3$ elliptic genus when the symmetry can be realized as a symmetry of some $K3$ sigma model.  
This proposal has passed a few consistency checks \cite{UMk3}, for instance for the group elements $g$ that can be realized as geometric symmetries.

In the latter category, we have learned a lot about the symmetries of $K3$ sigma models in the past years. First, a CFT analogue of Mukai's classification theorem of hyper-K\"ahler-preserving (or symplectic) automorphisms of $K3$ surfaces \cite{Mukai} has been established for $K3$ sigma models. Extending the lattice arguments in \cite{Kondo}, it was shown in \cite{GHV} that all symmetries of non-singular $K3$ CFTs preserving ${\cal N}=(4,4)$ superconformal symmetry are necessarily subgroups of the Conway group (${\rm Co}_0$, often known as the automorphism group of the Leech lattice) that moreover preserve at least a four-dimensional subspace in the irreducible 24-dimensional representation of the group. (Throughout the paper we will call such subgroups the ``4-plane preserving subgroups.")
This classification was later rephrased in terms of automorphisms of derived categories on $K3$ in \cite{HuybrechtsDerived}, and was moreover proposed to govern the symmetries of the appropriately defined moduli space relevant for $K3$ curve counting \cite{Cheng:2015kha}. 
Partially motivated by the proposed relation between umbral moonshine and $K3$ CFTs, this classification has been extended to include the singular CFTs in the moduli space of $K3$ sigma models \cite{Cheng:2016org}. Inspired by the classification in \cite{GHV,HuybrechtsDerived}, a fascinating conjecture was made in \cite{DuncanMackCrane} on the relation between these symmetries in the $K3$ setup and the Conway moonshine module  \cite{FLM,FLMBerk,Duncan,JohnSander}. The Conway moonshine module is a chiral superconformal field theory with  $c=12$ and symmetry given by the Conway group.
In particular, to each 4-plane preserving conjugacy class $[g]$ of the Conway group $\Co_0$ one attaches at most two weight 0 Jacobi forms, denoted  $\f_{\widehat g}$ and $\f_{\widehat g'}$ ($\f_{\widehat g}=\f_{\widehat g'}$ whenever the element $g$ fixes more than a 4-plane). 
In \cite{DuncanMackCrane} it was conjectured that $K3$ twining genera coincide with such functions  $\f_{\widehat g}$ and $\f_{\widehat g'}$ arising from the Conway moonshine module. 

One of the motivations of the current work is to gather evidence for (or against) the aforementioned proposals relating symmetries of $K3$ CFTs to umbral and Conway moonshine. One of the difficulties is that, apart from the special loci of torus orbifolds which describe theories  exhibiting  atypical symmetries  not directly related to umbral moonshine \cite{GHV}, we have extremely little computational control at generic points in the moduli space of  $K3$ sigma models. 
The Landau-Ginzburg (LG) orbifold description of $K3$ sigma models provides a powerful way in which symmetries can be studied explicitly and twining genera can be easily computed. 
Recall that LG orbifold theories are a 2d quantum field theory with ${\cal N}=(2,2)$ supersymmetry. In this paper we focus on such theories that flow in the IR to a $c=6,  ~{\cal N}=(4,4)$ superconformal field theories in the moduli space of $K3$ sigma models. The first goal of this article is to exploit the power of the UV description to gather more data on the  twining genera of $K3$ CFTs, and to compare them with the predictions from umbral and Conway moonshine. 
Needless to say, such data will be extremely valuable for  future endeavors to further elucidate the precise relation between umbral moonshine and $K3$ physics.

A second motivation to study the LG orbifold theories as models for symmetries of $K3$ CFTs is the following. Due to the topological nature of the elliptic genus, the computation of the twining genera depends only very roughly on the detailed properties of the theory. As a result, which we will explain in detail in \S\ref{subsec:The Symmetries of Landau-Ginzburg Orbifolds}, twining genera arising from symmetries of different models often share a uniform description. This leads to the possibility that the LG description furnishes a framework to combine symmetries arising from different points in the moduli space of the $K3$ CFT to obtain the action of a larger group (on the $\xoverline{Q}_+$ cohomology whose graded trace gives rise to the elliptic genus). We will explore this possibility with a specific example in \S\ref{sec:m11}. 

By exploiting the LG description of  $K3$ sigma models and studying their manifest discrete symmetries which are often inaccessible in the IR, we are able to realize certain predictions from umbral moonshine for the first time in a physical theory. Moreover, from these data we offer strong evidence for the relation between umbral moonshine and $K3$ CFTs. In particular, the results in this paper confirm that not only $M_{24}$ moonshine but also the other 22 cases of umbral moonshine appear to be relevant for symmetries of $K3$ CFTs, as we have also obtained twining genera coinciding with the predictions of umbral moonshine corresponding to the root systems $X=12A_2, 6D_4$ and $8A_3$. Next, by considering a novel type of ``asymmetric symmetries" we obtain twining genera coinciding with functions arising from umbral moonshine with complex multiplier systems, which have previously not been realized in the  context of $K3$ sigma models to the best of our knowledge, including those that do not arise from the Conway module. 
Finally, by exploiting the invariance of elliptic genus under $\xoverline{Q}_+$-exact deformations, we arrive at a uniform description of all twining functions corresponding to group elements of $M_{11}\subset 2.M_{12}$, as predicted by the $X=12A_2$ umbral moonshine, despite the fact that there is no LG model with symmetry group as large as $M_{11}$. 

  The plan of the rest of the paper is as follows. In \S \ref{sec:lgmm}, we  review the necessary aspects of the correspondence between Landau-Ginzburg theories and minimal models, and its generalization to orbifolds thereof. We  also review the computation of the elliptic genus in such theories, and the twining genera when the theory possesses a discrete symmetry. 
  
  In \S \ref{sec:data} we will study specific Landau-Ginzburg orbifolds describing $K3$ sigma models  and present their twining genera. Our primary examples will be  theories with six chiral superfields and cubic superpotentials, of which the most famous representative is the Fermat model which flows to the $(1)^6$ Gepner model.  Their  geometric interpretation is described in Appendix \ref{sec:K32}. 
Their ${\cal N}=(4, 4)$-preserving automorphism groups include certain maximal subgroups of umbral groups and give rise to interesting twining functions coinciding with  predictions from umbral and Conway moonshine. Analogously, in \S\ref{subsec:Quartics} we will analyze theories with four chiral superfields and quartic superpotentials. The explicit descriptions of  the symmetry groups of all the models discussed, as well as the tables recording the twining data, are collected in Appendix \ref{app:groupactions}.

 In \S\ref{sec:asymm} we explore a novel type of symmetry of  LG orbifold models -- those that act differently on different components of a chiral superfield  and preserve two of the supercharges and the Lagrangian of the theory.  Considering this class of symmetries allows us to recover more functions predicted by umbral moonshine. In particular, they often lead to twining genera with complex multiplier systems which have not been previously realized in the context of $K3$ CFTs. 
 In \S\ref{sec:m11}, we present a uniform description for the twining functions corresponding to group elements of $M_{11}\subset 2.M_{12}$, given by the natural 12-dimensional representation of the Mathieu group $M_{11}$. We conclude in \S\ref{sec:con} with a summary and discussions on open questions and future directions.

\section{LG Orbifolds, Gepner Models, and Geometry}\label{sec:lgmm}

In this section we will first briefly review the connection between $\mathcal{N}=(2,2)$ Landau-Ginzburg  models and $\mathcal{N}=(2, 2)$ superconformal minimal models (MM). In \S\ref{sec:LG Orbifolds, Gepner Models and Sigma Models} we will review the relation between LG orbifolds, Gepner models, and the sigma models describing Calabi--Yau manifolds. In particular, we summarize the computation of the elliptic genera of the corresponding CFTs from the point of view of LG orbifolds.
We will pay special attention to our main cases of interest:  LG orbifolds describing $c=6$, ${\cal N}=(4,4)$ superconformal theories lying in the moduli space of $K3$ sigma models. 
In \S\ref{subsec:The Symmetries of Landau-Ginzburg Orbifolds} we discuss specific types of symmetries of LG orbifolds and derive an expression for the corresponding twining genera.

\subsection{A Review of the $\mathcal{N}=2$ LG/MM Correspondence}

In this subsection we review the connection between $\mathcal{N}=2$ Landau-Ginzburg models and $\mathcal{N}=(2, 2)$ superconformal minimal models following \cite{Witten}\footnote{For more references containing early tests of the two dimensional LG/MM correspondence with various numbers of supercharges, see for example \cite{FendleyIntriligator, Cecotti, LercheVafaWarner, KastorMartinecShenker}.}.

An $\mathcal{N}=(2, 2)$ superconformal field theory, of which our $\mathcal{N}=(4, 4)$ models are a special case, is characterized by the following operator product expansions
\begin{subequations}
\begin{align}
%\begin{eqnarray*}
T(z) G^{\pm}(0) \, &\sim \,   \frac{3}{2 z^2}G^{\pm}(0) + \frac{1}{z} \partial G^{\pm}(0) \notag \\ 
T(z) J(0) \, &\sim \, \frac{1}{z^2}J(0) + \frac{1}{z} \partial J(0)\notag \\
G^+(z)G^-(0) \, &\sim \, \frac{2 c}{3 z^2} + \frac{2}{z^2}J(0) + \frac{2}{z}T(0) + \frac{1}{z}\partial J(0)\notag \\
G^+(z)G^+(0) \, &\sim \, G^-(z)G^-(0) \sim 0 \notag\\
J(z)G^{\pm}(0) \, &\sim \, \pm \frac{1}{z}G^{\pm}(0)\notag \\
J(z)J(0) \, &\sim \, \frac{c}{3 z^2}\notag~,
%\end{eqnarray*}
\end{align}
\end{subequations}
together with the corresponding right-moving counterparts. 
Here $T(z)$ is the stress-energy tensor, $J(z)$ is the $U(1)$ R-current, and $G^{\pm}(z)$ are the two supercharges. The $\mathcal{N}=2$ unitary minimal models form a discrete series with central charges given by $c= 3 \frac{k}{k + 2},\, k = 1, 2, ...$ . The minimal models also have a coset description that takes the form 
${\mathfrak{su(2)}_{k} \oplus \mathfrak{u(1)}_2 \over \mathfrak{u(1)}_{k + 2}}$.
 Importantly, the models (that moreover possess the spectral flow symmetry, see for instance \cite{Gray:2008je}) 
 enjoy an $ADE$ classification \cite{CIZ, Gepner:1986hr}  based on how one combines left- and right-moving characters of superconformal algebras to form modular invariant partition functions. In this language, $k+2$ is the Coxeter number of the relevant $ADE$ Dynkin diagram.
 This $ADE$ classification has an avatar in the LG picture, namely the $ADE$ classification of singularities, or catastrophes, that appear in the superpotential \cite{VafaWarner,Martinec:1988zu}.

While minimal models are genuine superconformal field theories, the LG models are generically  massive, super-renormalizable $\mathcal{N}=2$ supersymmetric quantum field theories. Such a theory has four supercharges, which come in two complex conjugate pairs labeled by their parity under the two-dimensional Lorentz group. They obey:
\begin{subequations}
\begin{align}
&Q_\pm^2=\xoverline Q_\pm^2=0, \qquad \{Q_\pm,\xoverline Q_\pm\}=2(H\mp P)\\
&\{\xoverline Q_+,\xoverline Q_-\}=2Z, \quad \, \{Q_+,Q_-\}=2Z^*\\
&\{Q_-,\xoverline Q_+\}=2\tilde Z,  \quad \,\{Q_+,\xoverline Q_-\}=2\tilde Z^*\\
&[F_V,Q_\pm]=-Q_\pm,  \quad [F_V,\xoverline Q_\pm]=\xoverline Q_\pm\\
&[F_A,Q_\pm]=\mp Q_\pm,  \quad [F_A,\xoverline Q_\pm]=\pm\xoverline Q_\pm
\end{align}
\end{subequations}
 where $Z, \tilde Z$ are central charges, and $F_V$ and $F_A$ generate two R-symmetries $U(1)_V$ and $U(1)_A$. We will be interested in theories with vanishing $Z, \tilde Z$ and conserved $U(1)_{V, A}$ symmetries.
 
 The supercharges can be represented as derivatives in superspace as
 \be\label{supercharges}
 Q_\pm=\frac{\partial}{\partial \theta^\pm} +i\xoverline \theta^\pm \left ( {\partial \over \partial x^0}\pm {\partial\over \partial x^1}\right ),~ \xoverline Q_\pm=-\frac{\partial}{\partial \xoverline\theta^\pm} -i \theta^\pm \left ( {\partial \over \partial x^0}\pm {\partial\over \partial x^1}\right ).
 \ee
 These commute with the superderivatives
  \be
 D_\pm=\frac{\partial}{\partial \theta^\pm} -i\xoverline \theta^\pm \left ( {\partial \over \partial x^0}\pm {\partial\over \partial x^1}\right ),~ \xoverline D_\pm=-\frac{\partial}{\partial \xoverline\theta^\pm} +i \theta^\pm \left ( {\partial \over \partial x^0}\pm {\partial\over \partial x^1}\right ).
 \ee
 which we can use to define certain superfields.
 
A chiral superfield is defined to obey
\be
\xoverline D_\pm \Phi=0
\ee
and can be expanded in components, suppressing the dependence on the worldsheet coordinates, as
\be
\Phi= \varphi + \sqrt 2 \theta^+\psi_+ + \sqrt 2 \theta^- \psi_- + 2 \theta^+\theta^- F + \ldots
\ee
where $\ldots$ contains derivative terms.

A Landau-Ginzburg Lagrangian is built out of these chiral superfields and takes the form
\begin{equation}\label{Lagrangian}
L = \int d^2 x \, d^4 \theta \,K(\Phi, \xoverline{\Phi}) - \int d^2 x \, d^2\theta\, W(\Phi) + h.c.
\end{equation} 
where $K(\Phi, \xoverline{\Phi})$ is the K{\"a}hler potential (D-term) which we will assume to be $\Phi \xoverline{\Phi}$ for simplicity and without loss of generality, and $W(\Phi)$ is the superpotential (F-term). We will always take the latter to be quasi-homogeneous in order to preserve the R-symmetry.  While the K{\"a}hler potential, which contains the kinetic terms, gets renormalized along the renormalization group (RG) flow, the superpotential does not. Thus, a LG theory may be considered to be completely characterized by its superpotential, at least when one is interested in RG-invariant quantities as we will be. 

One of Witten's tests of the LG/MM correspondence goes as follows. Recall that the elliptic genus of an ${\cal N}= (2,2)$ supersymmetric conformal field theory in the Ramond-Ramond sector is defined as \cite{SW1, SW2, SW3}
\begin{equation}
Z(\tau, z) = \textrm{Tr}_{{\cal H}_{RR}}\bigl((-1)^F y^{J_0} q^{L_0 -c/24} \xoverline{q}^{\bar{L}_0 - c/24}\bigr)
\end{equation}
where $q = e^{2 \pi i \tau}, y = e^{2 \pi i z}$, $F = F_L + F_R$ is the total fermion number, the sum of left and right-moving fermion numbers, and $J_0$ is the zero-mode of the left-moving $U(1)$ R-symmetry generator. The variables $(\tau, z)$ are valued in $\mathbb{H} \times \mathbb{C}$ and can be viewed as the chemical potentials for the energy and $U(1)$ charge of the theory, respectively. Moreover, the superconformal algebra dictates that $(-1)^{F_R} = (-1)^{\overline{J}_0}$ on all states and similarly for the left-movers.
We have placed Ramond boundary conditions on both the left and right-moving fermions. For a compact theory this is a holomorphic function, receiving only contributions from right-moving ground states. Moreover, it is actually a weak Jacobi form of weight 0 and index $d/2$ in the case of a SCFT with $c=3d$, in particular a sigma model for a Calabi-Yau $d$-fold \cite{KYY, Gritsenko}. Recall that a weak Jacobi form $\phi(\tau, z)$ of weight $k$ and index $m$ is a function satisfying the following modular and elliptic transformation laws:
\begin{align}\label{jac_form}
\phi\left(\frac{a \tau + b}{c \tau + d}, \frac{z}{c \tau + d}\right) & = (c \tau + d)^k e^{2 \pi i m \frac{c z^2}{c \tau + d}} \phi(\tau, z)\, , \  \bem a&b\\c&d\eem \in SL_2(\mathbb{Z}) \\
\phi \left(\tau, z + \m \tau + \l \right) & = e^{- 2 \pi i m (\m^2 \tau + 2 \m z)} \phi(\tau, z)\, , \qquad\quad \m,\l \in \mathbb{Z},
\end{align}
and is moreover bounded as $\tau \to i\infty$ for any given $z\in \CC$.
The elliptic genus has the advantage of being more readily computable than the full partition function of a theory, and is constant in a connected component of the CFT moduli space. Moreover, it is invariant under RG flow. This property will enable us to compute it directly in the (non-superconformal) LG theory in the UV.

Since the $\mathcal{N}=2$ minimal models are conformal field theories, their elliptic genus can be computed directly. First, one computes the characters of $\mathcal{N}=2$ superconformal algebra \cite{Dobrev, Kiritsis} in a representation $\mathcal{R}$:
\begin{equation}
\chi_{\mathcal{R}}(\tau, z) = \textrm{Tr}_{\mathcal{R}}(-1)^F q^{L_0 - c/24} y^{J_0}
\end{equation}
The representation $\mathcal{R}$ is labeled by the eigenvalues of the highest weight state under the left-moving Hamiltonian and $U(1)_R$ zero modes, $L_0, J_0$, respectively. 
Witten focuses on the A-series of minimal models, which means that combining left and right-movers into a full partition function means simply taking a diagonal sum of tensor products of the left and right-moving ${\cal N}=2$ multiplets.  After specializing the partition function to the elliptic genus (by turning off the right-moving $U(1)_R$ chemical potential), he obtains
\begin{equation}
Z(\tau, z) = \sum_{\alpha} {}^{\prime} \chi_{\mathcal{R}_\alpha}(\tau, z)
\end{equation} 
where the prime on the sum means to sum over representations $\mathcal{R}_{\alpha}$ whose ground state has  $H_R = 0$. In this case one can also express the characters themselves in terms of the elliptic genus.

To compare to the corresponding quantity in LG models, Witten computes the path integral of the latter on a torus with appropriate boundary conditions. The elliptic genus can be interpreted as the index of a certain supercharge in the LG theory, given by
\begin{equation}\label{eq:supercharge}
\xoverline{Q}_{+} = \int dx^1 \left(i \xoverline{\psi}_+ (\partial_0 + \partial_1)\varphi + \frac{\partial W}{\partial \varphi}\psi_{-} \right).
\end{equation}
Since the elliptic genus is an index, it is invariant under continuous deformations of the Lagrangian and one can turn off the superpotential and compute the elliptic genus in a free field limit. Here, the twisted boundary conditions of the elliptic genus, corresponding to the $y^{J_0}$ insertion, ensure that the path integral is still convergent in this limit \cite{Witten}.

An $A$-type superpotential for a single chiral superfield is simply $\frac{\Phi^{k+2}}{(k+2)}$, corresponding to the Dynkin diagram $A_{k+1}$.
In this case, the above considerations lead to the result
\be\label{AtypeEG}
Z_{k}(\t,z) = \frac{\th_1(\t, {k+1\over k+2} z)}{\th_1(\t, {1\over k+2} z)},
\ee
for the elliptic genus, where 
 \be
\th_1(\t,z)
	= -i q^{1/8} y^{1/2} \prod_{n=1}^\infty (1-q^n) (1-y q^n) (1-y^{-1} q^{n-1}). 
\ee

To understand the above formula, first we shall derive the $R$-charges of the various component fields of $\Phi$ and $\xoverline\Phi$. One finds the following SUSY transformations, written in components and after eliminating the auxiliary field by its equation of motion:
\begin{subequations}
\begin{align}
\label{eq:susy}
 \delta\varphi &= \sqrt{2} (-\epsilon_-\psi_+ + \epsilon_+ \psi_-)\\
 \delta\psi_+ &= i\sqrt 2(\partial_0 +\partial_1)\varphi\xoverline{\epsilon}_-+\sqrt 2 \epsilon_+\frac{\partial \xoverline{W}}{\partial \xoverline{\varphi}}\\
  \delta\psi_- &= - i\sqrt 2(\partial_0 -\partial_1)\varphi \xoverline{\epsilon}_++\sqrt 2 \epsilon_-\frac{\partial \xoverline{W}}{\partial \xoverline{\varphi}} \,.
\end{align}
\end{subequations}
 
	The Lagrangian should be invariant under the left-moving $U(1)$ transformation by whose charge $J_0$ one grades the elliptic genus. The supersymmetry generators $\epsilon^+=-\epsilon_-$ have charge 0 while $\epsilon^-=\epsilon_+$ have charge 1 under this $U(1)$. This in turn fixes the charges of all the component fields; we list them in Table \ref{tbl:Rcharges}.
\begin{table}[htb]
\begin{center}
\begin{tabular}{c|c}
Field&$J_0$ eigenvalue \\\midrule
$\varphi$& $w$ \\
$\psi_+$& $w$ \\
$\psi_-$& $-1+w$ \\
\end{tabular}\caption{\small R-charge assignments. We write $w=1/(k+2).$}\label{tbl:Rcharges}
\end{center}
\end{table}
Note that $\varphi$ and $\psi_+$ have positive charges whereas $\psi_-$ has negative charge. 
Subsequently, by treating $\varphi$, $\psi_+$ and $\psi_-$ as free fields we readily arrive at the answer \eq{AtypeEG} \cite{Witten}. 

The expectation that the elliptic genera computed using the minimal model and the LG model coincide can be turned around to give conjectural expressions for  the characters of the $\mathcal{N}=2$ minimal models in terms of the free-field formula arising from the LG model. 
This equality was subsequently checked for general $ADE$-type superpotentials in \cite{DiFY}.

Moreover, by studying the cohomology of $\xoverline{Q}_+$, Witten finds a chiral (purely left-moving) ${\cal N}=2$ superconformal algebra expressible in terms of the LG fields, in spite of the LG theory not being conformal itself. Since this algebra acts in cohomology, it also admits a free field representation: 
\begin{subequations}\label{eq:algebra}
\begin{align}
J_- &= {k+1\over k+2}\psi_-\xoverline \psi_--{i\over k+2}\varphi\vec\partial\bar\varphi\\
T_- &= 2\vec\partial\varphi\cdot\vec\partial \bar\varphi + i\bigl( \psi_-\vec\partial\bar\psi_--\vec\partial\psi_-\cdot\xoverline\psi_- \bigr)+{1\over k+2}\vec\partial\bigl( i\psi_-\xoverline\psi_--\varphi\vec\partial\xoverline\varphi \bigr)\
G_- &= -i\sqrt 2\psi_-\vec\partial\xoverline \varphi\\
\xoverline G_- &= i \sqrt 2\Bigl({k+1\over k+2}\Bigr)\vec\partial \varphi\cdot \xoverline \psi_- - {i \sqrt 2\over k+2}\varphi\vec\partial \bar\psi_-
\end{align}
\end{subequations}
where $\vec\partial=\partial_0-\partial_1$. This algebra has the same central charge, $c={3k\over k+2}$, as the corresponding $\mathcal{N}=2$ minimal model.

\subsection{LG Orbifolds, Gepner Models, and Sigma Models}
\label{sec:LG Orbifolds, Gepner Models and Sigma Models}

Though the central charge of a single minimal model is too small by itself to furnish a good candidate 
to describe a CY sigma model, one can consider tensor products of  minimal models. Indeed, Gepner \cite{Gepner} showed that this strategy, facilitated by an orbifold to project onto states with integral $U(1)$ charges, produces consistent and in principle exactly soluble string vacua. 
Through the LG/MM correspondence reviewed in the previous subsection, we thereby obtain a correspondence between LG orbifolds and Calabi-Yau sigma models. 
This correspondence has been studied extensively, for example in \cite{GreeneVafaWarner}, and put into a beautiful framework in terms of gauged linear sigma models (GLSMs) by Witten in \cite{Phases}.

In light of this correspondence and what we discussed before, we expect that the elliptic genus of a CY sigma model with a corresponding LG orbifold description to be computable via free fields. 
Indeed, the elliptic genera of LG orbifolds have been computed by Kawai, Yamada, and Yang in \cite{KYY}, who subsequently verified the matching with computations from the Gepner models. We will now summarize their results. 

Consider a $\grpZ{h}$ orbifold of a LG theory with $N$ chiral superfields of weights $\omega_i = 1/(k_i + 2)$, where the weights are defined by the transformation of the quasi-homogeneous superpotential:
\begin{equation}\label{quasi-hom_gen}
\lambda^w W(\Phi_1, \ldots, \Phi_N) = W(\lambda^{\omega_1} \Phi_1, \ldots, \lambda^{\omega_N}\Phi_N).
\end{equation}
The central charge of the theory is $\hat{c} \equiv c/3 = \sum_{i=1}^N \frac{k_i}{k_i + 2}$. 
In the multifield case, we can again define a supercharge $\xoverline{Q}_+$ and an  ${\cal N}=2$ superconformal algebra which acts on its cohomology. They are given by
\be \label{multifields_SCA}
\xoverline{Q}_+ = \sum_i \xoverline{Q}_{+,i}~,~~ J_- =  \sum_i  J_{-,i} ~,~~ T_- =  \sum_i T_{-,i} , 
\ee
and similarly for $G_-$ and $\xoverline G_-$, where the individual $i$-th components are given as in \eq{eq:supercharge} and  \eq{eq:algebra}.

The elliptic genus of the orbifolded theory is then 
\begin{equation}\label{eq:orbeg}
Z(\tau, z) = \frac{1}{h} \sum_{a, b \in \ZZ/h}(-1)^{\hat{c}(a + b + ab)} e^{2 \pi i (\hat{c}/2)(a^2 \tau + 2 a z)}\prod_{i=1}^N Z_{k_i}(\tau, z + a \tau + b)
 \end{equation}
 with $Z_{k_i}$  as in \eq{AtypeEG}.
 For a tensor product of $A_{k+1}$-type minimal models, we simply orbifold by $\grpZ{h}$, where $h=k+2$. More generally, $h = \textrm{lcm}(k_i + 2)$, where $k_i+2$ again coincides with the Coxeter number of the corresponding $i$-th Dynkin diagram.

In a more mathematical language, the above orbifold free-field expression for the elliptic genus can be understood through the following. Recall that in almost all cases, the Landau--Ginzburg model can be thought of as describing a CY sigma model with K{\"a}hler parameter taken to minus infinity \cite{Phases}. The corresponding CY hypersurface is parametrized by the chiral multiplets of the LG theory, when viewing the LG superpotential as defining a hypersurface in (weighted) projective space \footnote{The cubic models that we will discuss in \S\ref{subsec:cubic} are an important exception. See Appendix \ref{sec:K32} for the geometric interpretation of these cases.}. In this sense, the LG and CY descriptions can be thought of as two phases of the same theory, and give rise to IR conformal field theories which are in the same moduli space.
In  \cite{Gorbounov_Malikov} a spectral sequence converging to the cohomology of the chiral de Rham complex over a Calabi-Yau hypersurface was constructed, and its first term is given by a $bc\beta\gamma$ orbifold discussed in \cite{Borisov:1998dw} closely related to the free field limit of the LG orbifolds. As the graded supertrace of this $bc\beta\gamma$ orbifold is precisely given by (\ref{eq:orbeg}), while that of the graded trace of the cohomology of the chiral de Rham complex on a Calabi-Yau manifold yields nothing but its elliptic genus, the work of \cite{Gorbounov_Malikov} constitutes a mathematical proof of \eq{eq:orbeg}, where the LHS is defined to be the corresponding Calabi-Yau elliptic genus.

In this paper we are interested in LG orbifolds which flow in the IR to $K3$ sigma models, and these are theories with  $k_i$'s such that $c=6$. Recall that $\mathcal{N}= (2, 2)$ superconformal theories with $c=6$ (and integral $U(1)$ charges) always have enhanced  $(4, 4)$ superconformal symmetry \cite{Eguchi:1988vra}, and the only two options for such theories are $T^4$ sigma models and $K3$ sigma models.\footnote{Strictly speaking, this is not actually proven. We thank the referee for pointing this out. Probably one also needs to assume that the chiral algebra is not extended beyond (small) $\mathcal N=4$. Counterexamples to this could be interesting; for example in 2d CFTs theories with large $\mathcal N=4$ superconformal symmetry,  the elliptic genus vanishes. }
(See, for instance, \cite{Aspinwall1, Aspinwall2, AspinwallMorrison} and references therein for foundational work on string theory on $K3$ and \cite{NahmWendland} for analysis and proof regarding these properties of the moduli space of $\mathcal{N}= (4, 4)$ theories at $c=6$). The former class of theories has vanishing elliptic genus, so in practice one simply needs to check that our elliptic genus does not vanish, or in particular, that the Euler character $Z(\tau \rightarrow i\infty, z=0) = 24$, to verify that the theory  indeed lies in the moduli space of $K3$ sigma models. 
To avoid notational confusion, in what follows we will use the special moniker $\textbf{EG}(\tau, z)$ resp. $\textbf{EG}_g(\tau, z)$ for the main objects we are interested in in this paper -- elliptic genera of LG orbifolds describing $K3$ sigma models and the corresponding twining genera.

\subsection{The Symmetries of Landau-Ginzburg Orbifolds}
\label{subsec:The Symmetries of Landau-Ginzburg Orbifolds}

We are interested in studying the symmetries of these $K3$ Landau-Ginzburg orbifold theories and in computing the corresponding  ``twining genera". These are  defined as elliptic genera with an extra insertion of a symmetry generator $g$:
\begin{equation}
Z_g(\tau, z) = \textrm{Tr}_{{\cal H}_{RR}}\bigl(g\, (-1)^F q^{L_0 - c/24}\xoverline{q}^{\bar{L}_0 - c/24} y^{J_0}\bigr). 
\end{equation}
Clearly, for this trace to be well-defined, $g$ must commute with $L_0$,  $\bar L_0$ and $J_0$.  
For the definitions of $L_0, J_0$ in the LG theory acting on the $\xoverline{Q}_+$ cohomology, see (\ref{eq:algebra}). 

In this section, we will always consider symmetry generators $g$ that a.) are symmetries of some LG superpotential and b.) preserve the  $\mathcal{N}=2$ superconformal algebra of (the $\xoverline{Q}_+$ cohomology of) the UV theory (\ref{eq:algebra}) and its right-moving counterpart, c.) preserve all four charged chiral ring elements (in the Ramond-Ramond sector). 
Note that these criteria lead to a corresponding symmetry $g_{\rm IR}$ in the IR that preserves a copy of  $\mathcal{N}=(2,2)$  superconformal algebra, as well as the $NS$--$NS$ ground state and its images under the ${\cal N}=4$ spectral flow. Explicitly, these are the states responsible for the $2y$ and the $2y^{-1}$ terms in the elliptic genus.
This guarantees that $g_{\rm IR}$ preserves the full $\mathcal{N}=(4,4)$  superconformal algebra. Subsequently, the non-renormalization of the index of $\xoverline{Q}_+$ cohomology guarantees that the twining genera of $g_{\rm IR}$ can be computed in the UV in the way we shall describe shortly. As expected from general CFT arguments, they will be weak Jacobi forms of weight 0 and index 1 for the Hecke congruence subgroup 
$$\Gamma_0(\textrm{ord}(g)) \equiv \left\lbrace \bem a & b\\ c & d  \eem \in SL(2, \mathbb{Z})\vert c \equiv 0~~ \rm{mod}~{\rm{ord}(g)} \right\rbrace$$ of $SL(2, \mathbb{Z})$, possibly with a non-trivial multiplier giving an extra phase in the transformation \eq{jac_form}.
Later in \S\ref{sec:asymm} we will also discuss more general symmetries that do not satisfy all the above conditions which however lead to twining genera with the correct modular properties and interesting $\mathcal{N}=4$ decompositions.

Since our LG orbifold genera are computed in free field theory, it is  straightforward to obtain a similar expression for the twining genera. To start with, we will focus on twining genera coming from  automorphisms of a UV superpotential, such as phase rotations or permutations of the chiral superfields. 
In view of the above, computing  the elliptic genera twined by such symmetries involves  a simple adaptation of the calculation leading to the formula \eq{AtypeEG}. 

For simplicity of the derivation we will first focus on the case of a single chiral superfield $\Phi$ multiplied by a single phase $\alpha$. In terms of the component fields, all of them get multiplied by a single phase $\alpha$. 
 This is always automatically a symmetry of the full action since the K\"ahler potential is a function of $\Phi \xoverline \Phi$.\footnote{More general K{\"a}hler potentials will transform by a K{\"a}hler transformation of the form $K(\Phi, \xoverline{\Phi}) \rightarrow K(\Phi, \xoverline{\Phi}) + f(\Phi) + \xoverline{f}(\xoverline{\Phi})$, which still leaves invariant all observables of the theory.} 

 Now we can follow the steps in \cite{Witten} and compute the contribution of each superfield to the elliptic genus, with the generator $g$ of the symmetry inserted in the trace.
 First consider $\varphi$ and its complex conjugate $\xoverline{\varphi}$. This field will contribute
 \be\label{free_1}
 \frac{1}{1-\alpha y^{1\over k+2}} \frac{1}{1-\alpha^{-1} y^{-{1\over k+2}}}\prod_{n=1}^\infty\frac{1}{(1-\alpha y^{1\over k+2}q^n)(1-\alpha^{-1} y^{-{1\over k+2}}q^n)
(1-\alpha y^{1\over k+2}\xoverline q^n)
(1-\alpha^{-1} y^{-{1\over k+2}}\xoverline q^n)
}.
 \ee
 The right-moving fermion $\psi_+$ and its complex conjugate $\xoverline \psi_+$ will contribute
 \be
 (\alpha^{1\over 2} y^{1\over 2(k+2)}-\xoverline \alpha^{1\over 2} y^{-{1\over 2(k+2)}})\prod_{n=1}^\infty (1-\alpha y^{1\over k+2} \xoverline q^n)(1- \alpha^{-1} y^{-{1\over k+2}}\xoverline q^n)
 \ee
and the left-moving fermion $\psi_-$ and its complex conjugate will contribute
\be
 (\alpha^{1\over 2} y^{-{k+1\over 2(k+2)}}- \alpha^{-{1\over 2}} y^{{k+1\over 2(k+2)}})\prod_{n=1}^\infty (1-\alpha y^{-{k+1\over k+2}}  q^n)(1- \alpha^{-1} y^{{k+1\over k+2}} q^n)
 \ee
 Putting this together, one sees that under the symmetry $\Phi\to \alpha \Phi$, this superfield will contribute
 \be\label{eq:product}
  \frac{ \alpha y^{-{k\over 2(k+2)}}(1-\xoverline \alpha y^{{k+1\over k+2}})}{1-\alpha y^{1\over k+2}} \prod_{n=1}^\infty\frac{(1-\alpha y^{-{k+1\over k+2}}  q^n)(1- \alpha^{-1} y^{{k+1\over k+2}} q^n)}{(1-\alpha y^{1\over k+2}q^n)(1-\alpha^{-1} y^{-{1\over k+2}}q^n)}.
 \ee
 In terms of the standard Jacobi theta-function, 
this can be rewritten as
\be\label{twinedPhi}
Z_{k,\l}(\t,z)=\frac{\theta_1(\t, {k+1\over k+2}z- \lambda)}{\theta_1(\t,   {z\over k+2}+\lambda)}
\ee 
where we have written $\alpha= e^{2 \pi i \lambda}$.

Now it is straightforward to compute the elliptic genus of a LG orbifold with a superpotential of $N$ superfields, $\Phi_i$, $i=1\ldots N$, and twined under a symmetry of the superpotential which rotates each of the fields by some phase, $g: \Phi_i \mapsto e^{2\p i \l_i} \Phi_i$. This will just be
\begin{equation}\label{eq:orbegtwine}
%\textbf{EG}
Z_g(\tau, z) = \frac{1}{h} \sum_{a, b \in \ZZ/h}(-1)^{\hat{c}(a + b + ab)} e^{2 \pi i (\hat{c}/2)(a^2 \tau + 2 a z)}\prod_{i=1}^N Z_{k_i,\l_i}(\tau, z + a \tau + b)
 \end{equation}
 with $Z_{k_i,\l_i}$  as in \eq{twinedPhi} and $h$ is as in \eq{eq:orbeg}.

 In \S\ref{sec:data} and \S\ref{sec:moretwining} we will be mostly interested in theories with $N$ superfields with the same weights, which have symmetries of the form $W(\xunderline{\Phi}) = W(\xunderline{\alpha} \cdot \xunderline{\Phi})$, where $\underline\a$ is an $N\times N$ matrix in $SL(N, \mathbb{C})$. 
 By going to a diagonal basis we can always describe these symmetries in terms of phase rotations. In this basis  $\xunderline{\alpha}=\text{diag}(e^{2\p i \l_1},\dots,e^{2\p i \l_N})$ and one can directly apply \eq{eq:orbegtwine}.

\section{Some $K3$ Models and Their Symmetries }\label{sec:data}

In this section we will apply the general results of the previous section to specific LG orbifolds which flow to CFTs in the moduli space of $K3$ sigma models. We mainly focus on two types of examples: in \S\ref{subsec:cubic} we discuss theories with six chiral superfields and cubic superpotentials; in  \S\ref{subsec:Quartics}  we discuss theories with four chiral superfields and quartic superpotentials.
The geometric interpretation of the cubic models and the relation to the Hilbert scheme of two points on a $K3$ surface will be discussed in Appendix \ref{sec:K32}. 
The explicit descriptions of the symmetry groups of the models  discussed in this section as well as the corresponding twining genera are collected in Appendix \ref{app:groupactions}.

\subsection{The Cubic Theories}\label{subsec:cubic}

In this subsection we consider LG orbifold theories with six chiral superfields and a cubic superpotential.
From the requirement $\hat c = \sum_i {k_i\over k_i+2} =2$, we see that these theories have the property that they have the largest possible number of chiral superfields for a $K3$ LG model. As a result, one might expect them to be good starting points to investigate the symmetries of $K3$ models.

We will denote the chiral superfields by $\Phi_i, i=1\ldots 6$. Let's first consider a Fermat-type superpotential:
\be\label{fermat_superpl}
{W}(\xunderline{\Phi})=\sum_{i=1}^6 \Phi_i^3.
\ee
By the (orbifolded) LG/MM correspondence, this model flows to the so-called $(1)^6$ Gepner model in the IR. This is a $\grpZ3$ orbifold of the tensor product of 6 copies of the $k=1$ minimal model (and hence the notation  $(1)^6$), corresponding to the Dynkin diagram $A_2$. 
The symmetries of this model and its twining genera have already been studied directly in the Gepner picture in \cite{GHV};  we include this example here for the sake of completeness and to facilitate the comparison between the LG and Gepner approaches. 
In  \cite{GHV} it was shown that the group of symmetries preserving the $\mathcal{N}=(4, 4)$ superconformal algebra of this model is $3^4\!:\!A_6$.\footnote{Throughout this paper, our notation for groups mostly follows that of \cite{ATLAS}; in particular `$n$' is shorthand for the cyclic group of order $n$, `$:$' denotes the semidirect product $\rtimes$ and $G=A\cdot B$ denotes a group with normal subgroup $A$ such that $G/A=B$.  } This is one of the S-lattice subgroups of ${\rm Co}_0$ and is not a subgroup of any of the 23 umbral groups \cite{Curtis, ConwaySloane}. It has also been conjectured that the corresponding sigma model describes a $\grpZ3$ orbifold of $T^4$, with the appropriate $B$-field turned on \cite{Gaberdiel:2013nya}.

We can compute the twining of the elliptic genus under an even permutation of the six superfields, namely an element of  $A_6$, by diagonalizing the permutation matrix and subsequently using the formula \eq{eq:orbegtwine}. 
This leads to 
\be\label{twining_cubic}
\textbf{EG}^{c}_{g}(\tau, z)={1\over 3}\sum_{a,b\in \ZZ/3} q^{a^2}y^{2a}\prod_{i=1}^6 \frac{\theta_1\left (\t, {2\over 3}(z+a\t+b)-\l_i\right)}{\theta_1\left (\t,{1\over 3}(z+a\t+b)+\l_i\right)},
\ee
where $(e^{2\p i \l_1},\dots,e^{2\p i \l_6})$ are the eigenvalues of the symmetry $g$ acting on the six chiral super fields. An identical formula will hold for general symmetries in $3^4\!:\!A_6$ since we can always diagonalize such a symmetry. 

As we mentioned in \S\ref{subsec:The Symmetries of Landau-Ginzburg Orbifolds},  such a formula holds for the twining genera of any $K3$ cubic model, even when the superpotential differs from the Fermat one \eq{fermat_superpl} and the theory no longer flows to the $(1)^6$ Gepner model nor describes the $T^4/\ZZ_3$ orbifold model.  In view of this, we give the formula an extra label $^c$ for ``cubic."  

In what follows we discuss more general cubic models. We will focus on superpotentials satisfying the so-called transversality condition. 
 This means that the only solution to the set of equations
\be
\frac{ d{\cal W}^c(\underline\Phi)}{d\Phi_1}=\cdots=\frac{ d{\cal W}^c(\underline\Phi)}{d\Phi_6}=0
\ee
is when all $\Phi_i=0$. The transversality condition guarantees smoothness of the CFT, ensuring that there are no flat directions where a continuum of states could arise.

For a given superpotential, we consider symmetries satisfying certain conditions. 
A necessary condition for (\ref{twining_cubic}) to yield a sensible decomposition into $\mathcal N=4$ characters is that the permutation matrix has determinant one. Namely, we consider symmetries with eigenvalues satisfying $\sum_{i=1}^6 \l_i =0 \xmod{\ZZ}$. This condition can be understood from  the requirement that the element 
\be\label{prod_is1}
\text{det}_{ij}\left( \frac{\pa^2 {\cal W}^c}{\pa \Phi_i\pa \Phi_j}\right) \Big\lvert_{\partial W\sim0} \sim \Phi_1 \Phi_2\Phi_3\Phi_4\Phi_5\Phi_6 
\ee
 of the chiral ring with the highest $R$-charge should remain invariant under the symmetry.

In the rest of this subsection we will make extensive use of the following recent result. An exhaustive search of symmetry groups
acting on transverse cubic equations with six variables was performed in \cite{HohnMason} in the context of classifying symplectic automorphisms of manifolds deformation equivalent to the Hilbert scheme of two points on a $K3$ surface, $K3^{[2]}$. The somewhat curious fact that a cubic equation in $\PP^5$ defines a 4-(complex)dimensional variety while the corresponding LG model describes a $K3$ sigma model is reflected in relations between cubic fourfolds, geometric symmetries of $K3^{[2]}$, and symmetries of $K3$ sigma models. This is a topic currently under active investigation in the realm of algebraic geometry  and we will summarize a part of this connection in Appendix \ref{sec:K32}.

In \cite{HohnMason}, 15 maximal symmetry groups have been identified, in the sense that any hyper-K\"ahler-preserving symmetry group of any manifold that is deformation equivalent to $K3^{[2]}$ is a subgroup of one of these 15 groups. Among them, explicit cubic equations baring the symmetry of six of the  groups have been identified. They are listed in Table \ref{tbl:groups_superpotentials} (cf. Table 11 of \cite{HohnMason}), where in $\mathcal{W}_4^c$ we have defined the following deformation:
\begin{multline}
\lambda.\sigma_3(\xunderline{\Phi}) =  {1 \over 5}(-3 \zeta_{24}^7 - 3 \zeta_{24}^5 + 3 \zeta_{24}^4 - 3 \zeta_{24}^3 + 6 \zeta_{24} - 3)\times  \\ \bigl( \Phi_1 \Phi_2 \Phi_3 + \Phi_1 \Phi_2 \Phi_4 + (\zeta_{24}^4 - 1) \Phi_1 \Phi_2 \Phi_5 + \Phi_1 \Phi_2 \Phi_6 + (\zeta_{24}^4 - 1)\Phi_1 \Phi_3 \Phi_4 + \Phi_1 \Phi_3 \Phi_5 + \Phi_1 \Phi_3 \Phi_6 + \\ (\zeta_{24}^4 - 1) \Phi_1 \Phi_4 \Phi_5 - \zeta_{24}^4 \Phi_1 \Phi_4 \Phi_6 - \zeta_{24}^5 \Phi_1 \Phi_5 \Phi_6 + (\zeta_{24}^4-1) \Phi_2 \Phi_3 \Phi_4 + (\zeta_{24}^4-1) \Phi_2 \Phi_3 \Phi_5 \\ - \zeta_{24}^4 \Phi_2 \Phi_3 \Phi_6 + \Phi_2 \Phi_4 \Phi_5 + \Phi_2 \Phi_4 \Phi_6 - \zeta_{24}^4 \Phi_2 \Phi_5 \Phi_6 + \Phi_3 \Phi_4 \Phi_5 - \zeta_{24}^4 \Phi_3 \Phi_4 \Phi_6 + \Phi_3 \Phi_5 \Phi_6 + \Phi_4 \Phi_5 \Phi_6 \bigr) \nonumber.
\end{multline}
Here and everywhere else in this paper we write $\zeta_n=e^{2\p i/n}$. An explicit description of the group action on each of the six superpotentials is given in Appendix  \ref{app:groupactions}.

\begin{table}[htb]
\begin{center}
\begin{tabular}{CCCC}\toprule
i&\text{Group  } G_i& \text{Root Systems } X &\text{Superpotential  } \mathcal{W}^c_i(\xunderline{\Phi}) \\\midrule
1&L_2(11)  & 12A_2 & 
  \Phi_0^3 + \Phi_1^2 \Phi_5 + \Phi_2^2 \Phi_4 + \Phi_3^2 \Phi_2 + \Phi_4^2 \Phi_1 + \Phi_5^2 \Phi_3 \\ \nonumber
2&(3 \times A_5):2  &6D_4& 
  \Phi_0^2 \Phi_1 + \Phi_1^2 \Phi_2 + \Phi_2^2 \Phi_3 + \Phi_3^2 \Phi_0 + \Phi_4^3 + \Phi_5^3 \\ \nonumber
3&A_{7}  &24A_1&  
\sum_{i=0}^5 \Phi_i^3 - (\sum_{i=0}^5 \Phi_i)^3 \\ \nonumber
4&M_{10}  &24A_1, 12A_2&
 \sum_{i=0}^5 \Phi_i^3 + \lambda.\sigma_3(\Phi_0, \ldots, \Phi_5)\\ \nonumber
5&3^{1+4}\!:\!2.2^2  &-& 
 \sum_{i=0}^5 \Phi_i^3 + 3(i - 2e^{\pi i/6} - 1)(\Phi_0 \Phi_1 \Phi_2 + \Phi_3 \Phi_4 \Phi_5) \\ \nonumber
6&3^4\!:\!A_6  &-& 
\sum_{i=0}^5 \Phi_i^3\\ \bottomrule
\end{tabular}\caption{\small The six maximally symmetric cubic superpotentials, their symmetry groups, and the associated cases of umbral moonshine.}\label{tbl:groups_superpotentials}
\end{center}
\end{table} 

Now it is a straightforward task to apply \eq{twining_cubic} to compute the corresponding twining genera.
We collect the results for all conjugacy classes $[g]$ of each group
in Tables \ref{tbl:cubicW1}-\ref{tbl:cubicW6}. 
For convenience we also summarize the data in Tables \ref{tbl:twinings} and \ref{tbl:Conway}.
We shall explain our notation shortly. 

We characterize the twinings by their order and the associated 24-dimensional Frame shape. 
Recall that, for $G$ a finite group and an $n$-dimensional representation $\r: G \to GL(n,\CC)$ of $G$ such that the corresponding characters are rational numbers, the Frame shape $\Pi_g$ of  an element $g$ of $G$  can be viewed as a convenient tool to label the eigenvalues of $g$. Suppose the eigenvalues of $\r(g)$ are given by $\a_1, \dots, \a_n$, then we say that  
 $$\Pi_g = \prod_{\substack{k\in \ZZ_{>0}\\ a_k \in \ZZ_{\neq 0}}} k^{a_k}$$ is the Frame shape if
 $$
 \prod_{k} (1-t^{k})^{a_k}  = \text{det}({\bf 1}-t \r(g) ).
 $$
The eigenvalues are then given by $\text{det}({\bf 1}-t \r(g) )= \prod_{i=1}^n (1-\a_i t)$.  

In general, the 24-dimensional Frame shape alone is not sufficient to determine the twining function. This is because two elements of $O(4,20;\ZZ)$ might have the same eigenvalues but  not belong to the same conjugacy class, and as a result there is no reason for them to lead to the same twining genus. An explicit example of this is the two conjugacy classes with the same order 12 Frame shape in the symmetry group $3^{1+4}\!:\!2.2^2$ of the LG model with superpotential $\mathcal{W}^c_5$ in Table \ref{tbl:groups_superpotentials}, which lead to two different twining genera  (cf. Table \ref{tbl:Conway} and Table \ref{tbl:cubicW6}). In the context of the conjectural relation between umbral moonshine and $K3$ twining genera \cite{UMk3}, a related fact is that elements of different umbral groups with the same Frame shape can be associated to different candidate twining genera, for instance there exist $g\in G({24A_1})$ and $g'\in G({12A_2})$ with $\Pi_g=\Pi_{g'} \!=\! 3^8$ that have $\phi^{24A_1}_g \neq \phi^{12A_2}_{g'}$. This behavior is to be contrasted with  twining functions corresponding to geometric symmetries, which are identical across all the umbral groups as is required by the consistency with the Torelli theorem.

Comparing with the predictions from umbral  \cite{UMk3} and Conway moonshine \cite{DuncanMackCrane}, we find the following results. 
First, all twining genera arising from the models with superpotentials ${\cal W}^c_1$, ${\cal W}^c_2$, ${\cal W}^c_3$, ${\cal W}^c_4$ coincide with predictions arising from certain instances of umbral moonshine. Second, for any of these four models, there exists at least one instance of umbral moonshine that captures all the twining genera with respect to all elements of the corresponding symmetry group. Concretely, for any $i\in \{1,2,3,4\}$ 
there exists at least one Niemeier lattice $N(X)$, labeled by its root system $X$, such that the following two conditions are satisfied: 1. $G_i\subset G(X)$, 2. ${\textbf{EG}}^X_g$ for all $g\in G_i$ coincide with  $\f^X_{g'}$ with the corresponding $g'$ given by the embedding of the group.  
In Table \ref{tbl:groups_superpotentials} we have listed the corresponding $X$ for each of these four superpotentials. 
Third, for ${\cal W}^c_1$, ${\cal W}^c_2$ and ${\cal W}^c_3$ such an assignment of a case of umbral moonshine is in fact unique. 
For instance,  the order 11 twining function coming from $G_1\cong L_2(11)$ arises only from umbral moonshine with $X=12A_2$, and the only case that accommodates  $G_3\cong A_7$ in the corresponding umbral group is when $X=24A_1$. 
Note also that for ${\cal W}^c_2$, the symmetry group $(3 \times A_5)\!:\!2 \subset 3.S_6$ is in fact a maximal subgroup. 

We also remark that the unique assignment of an $X$ to our models ${\cal W}^c_{1,2,3}$ is consistent with the so-called discriminant property of  umbral moonshine, in the sense that the number field generated by the characters of $G_i$ is contained in that generated by the characters of $G(X)$ in a non-trivial way in all the three cases.
In light of this, it is tempting to associate the fourth theory to the case of $X=12A_2$, although both $12A_2$ and $24A_1$ accommodate the symmetries and the corresponding twining functions of the  ${\cal W}^c_4$ model. This is because  the number field generated by the characters of $G_4\cong M_{10}$ is contained in that generated by the characters of $G(12A_2)\cong 2.M_{12}$ but not in that  of $G(24A_1) \cong M_{24}$.

Notably, not only the case with $X=24A_1$,  also often referred to as Mathieu moonshine, shows up in our analysis. 
Rather, the cases with $X=12A_2$ and $X=6D_4$ play an equally prominent role in describing the symmetries of the models we analysed. 
This can be viewed as supporting evidence for the idea that all 23 cases umbral moonshine are relevant for the symmetries of $K3$ string theory. 
Finally, in all six cases studied in this subsection, the twinings can be obtained from the Conway module of \cite{DuncanMackCrane}, lending non-trivial support to the conjectural relevance of this Conway module for the ${\cal N}=(4,4)$-preserving symmetries of $K3$ sigma models. (See \S\ref{sec:asymm} for twining genera that do not arise from Conway moonshine.)

An obvious question arising here is the following. 
When and under what conditions can/should a LG orbifold with a given superpotential (and subsequently the corresponding sigma model in the IR)  be associated to a given instance of umbral moonshine in the sense discussed above? 
We will postpone this discussion until \S\ref{sec:con}. 

In view of the above, for the first four cases we specify the twining functions by giving the corresponding $\phi^{X}_{g}$ arising from umbral moonshine in Appendix \ref{subsec: twining tables}, where $[g]$ is denoted using the standard Atlas label for the conjugacy classes of the corresponding umbral group $G(X)$ (see also \cite{umbraltwo,UMk3}).
 In the cases of ${\cal W}^c_1$ , ${\cal W}^c_2$, ${\cal W}^c_3$ we have a unique such choice of $X$ while in the case of ${\cal W}^c_4$ we choose to use $X=12A_2$ for our labeling.  
 Similarly, we split the summary of all twinings arising from these cubic models amongst the two Tables \ref{tbl:twinings} and \ref{tbl:Conway} according to whether or not they are  associated to an umbral group. 
 In Table \ref{tbl:twinings}, all but two functions can arise as geometric symmetries and the corresponding twining genera are uniquely specified by the 24-dimensional Frame shape.  We use the standard $M_{24}$ labels for the conjugacy classes in these cases. If the twining genera are uniquely associated to a particular umbral group, we append the label $X$ of the root system of the corresponding Niemeier lattice. In Table \ref{tbl:Conway}, we use hatted names to label the given function. The first few terms of the Fourier expansions of these twining genera are also included in Appendix \ref{sec:conwaytwinings} for the convenience of the reader. In the final column, we list which superpotentials of those in Table \ref{tbl:groups_superpotentials} yield this twining function.

\begin{table}[htb]
\begin{center}
\begin{tabular}{c|c|c|c|c}
Order of $g$& $\Pi_g$ & $\textbf{EG}^c_g$& $X$&$\mathcal W_i^c$\\\hline
1& $1^{24}$& $\phi_{1A}^X$&$24A_1$&$1,2,3,4,5,6$\\
2& $1^82^8$& $\phi_{2A}^X$&$24A_1$&$1,2,3,4,5,6$\\
3 & $1^63^6$&$\phi_{3A}^X$&$24A_1$&$1,2,3,4,5,6$\\
4&$1^42^24^4$& $\phi_{4B}^X$&$24A_1$&$2,3,4,5,6$\\
5& $1^45^4$&$\phi_{5A}^X$&$24A_1$&$1,2,3,4,6$\\
6& $1^22^23^26^2$&$\phi_{6A}^X$&$24A_1$&$1,2,3,6$\\
7& $1^37^3$&$\phi_{7A}^X$&$24A_1$&3\\
8&$1^22.4.8^2$& $\phi_{8A}^X$&$24A_1$&4\\
11& $1^211^2$&$\phi^X_{11AB}$&$12 A_2$&1\\
15& $1.3.5.15$&$\phi^X_{15AB}$&$6 D_4$&2\\
\end{tabular}\caption{\small Twining genera in the cubic LG models which can arise from umbral moonshine.  
 }\label{tbl:twinings}

\vspace{7pt}
\begin{tabular}{c|c|c|c}
Order of $g$& $\Pi_g$  & $\textbf{EG}^c_g$&$\mathcal W_i^c$\\\hline
3&$3^9/1^3$&$\phi_{\widehat{3C}}$&$5,6$\\
6&$1^53.6^4/2^4$&$\phi_{\widehat{6I}}$&$5,6$\\
9& $1^3 9^3/3^2$&$\phi_{\widehat{9C}}$&6\\
9&$1^3 9^3/3^2$&$\phi_{\widehat{9C'}}$&6\\
12&$1.2^2 3.12^2/4^2$&$\phi_{\widehat{12L}}$&5\\
12&$1.2^2 3.12^2/4^2$&$\phi_{\widehat{12L'}}$&5\\
\end{tabular}\caption{\small Twining genera in the cubic LG models which do not arise from umbral moonshine. See Appendix \ref{sec:conwaytwinings} for the details of these functions.
 }\label{tbl:Conway}
\end{center}
\end{table}

Before we close this subsection, we comment that the twining genera listed in Table \ref{tbl:twinings} and \ref{tbl:Conway} often arise in other models with superpotentials different from the maximally symmetric ones given in Table \ref{tbl:groups_superpotentials}. 
If one is trawling the moduli space of some set of LG theories in search of high-order symmetries, one will in general not have the benefit of a classification \`a la \cite{HohnMason}.
Fortunately, other methods exist to find particularly symmetric points in the LG moduli space. One such trick is to start with the Fermat superpotential (\ref{fermat_superpl}) and consider various nonlinear field redefinitions, plus an additional orbifoldization to ensure single-valuedness of the chiral superfields, as in \cite{GreenePlesser}. Some examples of superpotentials with interesting symmetries obtained by this trick include:  
\be
W(\underline \Phi)=\Phi_1^2\Phi_2 + \Phi_2^2 \Phi_3 + \Phi_3^2\Phi_4 +\Phi_4^2\Phi_5 + \Phi_5^2 \Phi_1 + \Phi_6^3.
\ee
One can check that this superpotential has the symmetry $W(\underline \Phi)=W(\underline \alpha \cdot \underline \Phi)$, which  diagonally rotates the chiral superfields by $\underline \alpha= \text{diag}( \zeta_{11},\zeta_{11}^9,\zeta_{11}^4,\zeta_{11}^3,\zeta_{11}^5,1)$.
Another example is 
\item \be
W(\underline \Phi)=\Phi_1^2\Phi_2 + \Phi_2^2 \Phi_3 + \Phi_3^2\Phi_4 +\Phi_4^2\Phi_5 + \Phi_5^2 \Phi_6 + \Phi_6^2\Phi_1
\ee
which admits an order 7 symmetry rotating the fields by  $\underline \alpha= \text{diag}(\zeta_{7},\zeta_{7}^5,\zeta_{7}^4,\zeta_{7}^6,\zeta_{7}^2,\zeta_{7}^3)$ and leading to the order 7 twining genus listed in Table \ref{tbl:twinings}.

\subsection{The Quartic Theories}
\label{subsec:Quartics}

In this subsection we will explore the symmetries of other LG superpotentials. In this case, the geometric interpretation is more straightforward: the LG superpotential $W^q(\xunderline{\Phi})$ gives the equation $W^q(\xunderline{X})=0$ defining a $K3$ hypersurface in $\mathbb P^3$.

\begin{table}[htb]
\begin{center}
\begin{tabular}{CCCC}\toprule
i&\text{Group  } G_i& \text{Root Systems }X &\text{Superpotential  } \mathcal{W}^q_i(\xunderline{\Phi}) \\\midrule
1&L_2(7)\times 2  & 8A_3 & 
\Phi_1^3 \Phi_2 + \Phi_2^3 \Phi_3 + \Phi_3^3 \Phi_1 + \Phi_4^4 
 \\  \nonumber
2&M_{20} &24A_1& \Phi_1^4 + \Phi_2^4 + \Phi_3^4 + \Phi_4^4 + 12 \Phi_1 \Phi_2 \Phi_3 \Phi_4 \\ \nonumber
3&T_{192}  &24A_1,  8A_3&\Phi_1^4 + \Phi_2^4 + \Phi_3^4 + \Phi_4^4 -2 i \sqrt{3} \left(\Phi_1 \Phi_2  + \Phi_3 \Phi_4 \right) \\ \nonumber
4&(2 \times 4^2)\!:\!S_4 &-&   \Phi_1^4 + \Phi_2^4 + \Phi_3^4 + \Phi_4^4 
 \\ 
\bottomrule
\end{tabular}\caption{\small The four quartic superpotentials, their symmetry groups, and the associated cases of umbral moonshine.}\label{tbl:groups_superpotentials_quartic}
\end{center}
\end{table} 

In this subsection we will investigate  four particularly symmetric $K3$ surfaces (or, equivalently, LG superpotentials),  given by quartic equations $ \mathcal{W}^q_i(\xunderline{\Phi})=0$ in $\mathbb{P}^3$, where $\mathcal{W}^q_1, \dots ,\mathcal{W}^q_4$ are listed in 
Table \ref{tbl:groups_superpotentials_quartic}.\footnote{Unlike the case of cubic superpotentials in six variables, the set of quartic superpotentials we discuss in this section does not necessarily furnish an exhaustive list of 
the largest symmetry groups acting on four-variable quartic superpotentials. Quartics in four-variables have a geometric interpretation as algebraic $K3$ surfaces and
 the particular superpotentials in Table 5 were identified by Mukai \cite{Mukai} in his classification of geometric symmetries; in some of these cases we 
 have identified non-geometric extensions of the geometric symmetry groups (e.g. the Fermat quartic viewed as a Gepner model has $(2 \times 4^2):S_4$
 symmetry, while symplectic automorphisms of a geometric K3 at the Fermat point has symmetry group $4^2:S_4$).}
 As we explained before, all the four-variable quartic models share the following expression for the elliptic genus, this time with a $\grpZ4$ orbifold action (cf. \eq{eq:orbegtwine}):
\begin{equation}
\textbf{EG}^q(\tau, z) = \frac{1}{4}\sum_{a, b \in \ZZ/4}q^{a^2}y^{2 a}  Z_{2,0}^4(\tau, z + a \tau + b). 
\end{equation}
Similarly, the twining genus associated to a symmetry rotating the four superfields by a phase $g: \Phi_i \mapsto e^{2\p i \l_i} \Phi_i$ is given by 
\be\label{twining_quartic}
\textbf{EG}^{q}_{g}(\tau, z)={1\over 4}\sum_{a,b\in \ZZ/4} q^{a^2}y^{2a}\prod_{i=1}^4 \frac{\theta_1\left (\t, {3\over 4}(z+a\t+b)-\l_i\right)}{\theta_1\left (\t,{1\over 4}(z+a\t+b)+\l_i\right)},
\ee
The twining genera corresponding to each of the four LG orbifolds listed in Table \ref{tbl:groups_superpotentials_quartic} are listed in Appendix \ref{app:groupactions} and summarized in Table \ref{tbl:quartictwinings}.

The  quartic ${\cal W}^q_1=0$, an extension of  Klein's famous quartic surface, has symmetry group $L_2(7) \times 2$ \cite{KleinBook}, a maximal 4-plane preserving subgroup of $Co_0$\footnote{For a more precise characterization of the maximal 4-plane preserving subgroups of $Co_0$, i.e. those groups that act as $\mathcal{N}=(4, 4)$-preserving symmetries in $K3$ sigma models, see \cite{GHV}. 
See \cite{HohnMason} for the subset of those groups that act as symplectic automorphisms of manifolds of $K3^{[2]}$-type in relation to the discussion in Appendix \ref{sec:K32}.
}. 
The $\ZZ_2$ is generated by the action which maps $\Phi_4$ to $-\Phi_4$ and the rest of $\Phi_i$'s invariant. Note that though this action is not symplectic in the geometric language, it preserves the $\mathcal N=(2,2)$ superconformal algebra and the top chiral ring element
\be\label{quartic_top_chiralring} 
\text{det}_{ij}\left( \frac{\pa^2 W}{\pa \Phi_i\pa \Phi_j}\right)\Big\vert_{\partial W \sim 0} \sim 
\Phi_1^2\Phi_2^2 \Phi_3^2 \Phi_4^2, 
\ee
as well as, by explicit computation, the rest of the charged RR ground states. Therefore, this extra order 2 symmetry preserves the full $\mathcal N=(4,4)$ algebra of equation (\ref{eq:algebra}). Here we have chosen the boson and fermions in $\Phi_4$ to transform as
\be
g: \varphi \mapsto e^{\pi i}\varphi, \psi_+ \mapsto e^{\pi i}\psi_+, \psi_- \mapsto e^{3 \pi i} \psi_-
\ee
in order to preserve of the NS vacuum. 
Note that $\mathcal W^q_1$ actually possesses more symmetries, for instance an order 4 symmetry of $\mathcal W^q_1$ which acts as 
\be
\Phi_4\mapsto \pm i \Phi_4. 
\ee
However, the former clearly fails the requirement of preserving the top chiral ring element \eq{quartic_top_chiralring}.

Another particularly symmetric quartic is given by $\mathcal{W}_2^q=0$, which, when viewed as geometric $K3$ surface, has symplectic automorphism group $M_{20} \simeq 2^4.A_5$ \cite{Mukai}; this group potentially gets extended in the LG orbifold phase, but we have not investigated this possibility. Note that $M_{24}\cong G(24A_1)$ is the only umbral group that can accommodate $M_{20}$.

The quartic ${\cal W}^q_3=0$ has symplectic automorphism group $T_{192}$\cite{Mukai}, which is a subgroup of both $M_{24}$ and $2.AGL_3(2)$. We have not explored if it admits a non-geometric extension in the LG phase though this would be an interesting possibility.\footnote{We do know from the work \cite{Frantzen} that unlike $\mathcal{W}_{1}^q$, neither $\mathcal{W}_{2}^q$ nor $\mathcal{W}_{3}^q$ admit any antisymplectic extensions of the form $G \times 2$, where $G=\left\lbrace T_{192}, M_{20} \right\rbrace$, but we have not explored the case of more intricate non-symplectic extensions.}

Finally we now turn to ${\cal W}^q_4$, the Fermat quartic. This Landau--Ginzurg orbifold flows in the IR to the Gepner model $(2)^4$, whose symmetries have been studied in \cite{GHV}, where the ${\cal N}=(4, 4)$-preserving symmetries of the $(2)^4$ model is found to be $(2 \times 4^2)\!:\!S_4$.

\begin{table}[htb]
\begin{center}
\begin{tabular}{c|c|c|c|c}
Order of $g$& $\Pi_g$ & $\textbf{EG}^q_g$& $X$&$\mathcal W^q$\\\hline
1& $1^{24}$& $\phi_{1A}^X$&$24A_1$&$1, 2, 3, 4$\\
2& $1^82^8$& $\phi_{2A}^X$&$24A_1$&$1, 2, 3, 4$\\
2 & $2^{12}$ & $\phi_{2B}^X$& $24A_1$ & $4$\\ 
3&$1^63^6$ &$\phi_{3A}^X$&$24A_1$&$1, 2, 3, 4$\\
4&$1^42^24^4$& $\phi_{4B}^X$&$24A_1$&$1, 2, 3, 4$\\
6&$1^2 2^2 3^2 6^2$&$\phi_{6A}^X$&$24A_1$&$1, 3, 4$\\
7&$1^37^3$&$\phi_{7AB}^X$&$24A_1$&$1$\\
8&$1^2 2.4.8^2$&$\phi_{8A}^X$&$24A_1$&$4$\\
5&$1^4 5^4$&$\phi_{5A}^X$&$24A_1$&$2$ \\
4&  $2^6 4^4/1^4$ & $\phi_{\widehat{4F}}$& $-$&$4$\\ 
14&$1.2.7.14$&$\phi^X_{14AB}$&$8A_3$&$1$\\
\end{tabular}\caption{\small Twining genera arising from the quartic LG models. We use the same notation as in Tables \ref{tbl:twinings} and \ref{tbl:Conway}.
}\label{tbl:quartictwinings}
\end{center}
\end{table}

Comparing these results with the predictions from umbral and Conway moonshine, we find conclusions very similar to  the cubic case. 
First, all twining genera arising from the models with superpotentials ${\cal W}^q_1$, ${\cal W}^q_2$, ${\cal W}^q_3$ coincide with predictions arising from certain instances of umbral moonshine. Second, for any of these three models, there exists at least one instance of umbral moonshine capturing all the twining genera. In Table \ref{tbl:groups_superpotentials_quartic} we have listed the corresponding $X$ for each of these four superpotentials. 
Third, for ${\cal W}^q_1$ and ${\cal W}^q_2$  such an assignment of a case of umbral moonshine is in fact unique. 
For instance, the order 14 symmetry of ${\cal W}^q_1$ leads to a twining genus only appearing in the $8A_3$ case of umbral moonshine. Finally, in all four cases the twinings can be obtained from the Conway module. In particular, again we see that the torus orbifold model, equivalent to the LG orbifold model describing the Fermat quartic, leads to twining genera that can only be captured by the Conway module.

\section{More on Twining Genera}\label{sec:moretwining}

In \S\ref{sec:asymm} we discuss more general classes of symmetries than those discussed in \S\ref{sec:data}. 
After explaining why twining genera can be defined for these more general symmetries, we study this type of symmetries explicitly in a few LG orbifold theories and note that they produce novel twining functions arising from umbral moonshine, but not always from Conway moonshine. 
In \S\ref{sec:m11}, inspired by the possibility of combining symmetries from different UV models we discuss how all twinings of the Mathieu group $M_{11}$  arising from $X=12A_2$ umbral moonshine admit a uniform description in terms of the natural 12-dimensional representation, although there is no UV model with  $M_{11}$ as the symmetry group.

\subsection{The Asymmetric Symmetries of LG Orbifolds}\label{sec:asymm}

In \S\ref{sec:data} we  studied  symmetries of LG orbifolds which (in some basis) rotate all components of a superfield by the same phase and  
manifestly preserve an obvious copy of the UV ${\cal N}=2$ superconformal algebra which acts in cohomology (cf. \eq{eq:algebra} and \eq{multifields_SCA}).
However, as we discussed above, in order to define a twining genus we only need to require that the symmetry of the SCFT  leave invariant (the zero modes of) the left- and right-moving energy momentum tensor, as well as the left-moving $U(1)$ current. 
In the UV language, as the elliptic genus computes the graded index of the cohomology of the supercharge $\xoverline{Q}_+$ \eq{eq:supercharge}, the twining genus is well-defined as long as the symmetry generator preserves the supercharge $\xoverline{Q}_+$.\footnote{Note that all symmetries we consider manifestly commute with the Hamiltonian $H_L$ and the left-moving $U(1)$ R-current $J_L$ which provide the grading.  }

In this subsection we will consider what we call the ``asymmetric symmetries": symmetries of the LG orbifold that preserve the supercharge $\xoverline{Q}_+$ and transform the different components of the chiral superfields in different ways. 
Perhaps somewhat surprisingly, we find that such symmetries very often lead to twining genera that coincide with  functions arising from umbral moonshine. It would be interesting to understand the detailed description of such asymmetric symmetries from the point of view of the IR CFT. 

Consider a symmetry acting on the components of the $i$-th superfield in a LG orbifold as 
$$g: \varphi_i \mapsto \a_i \varphi_i, ~\psi_{+,i} \mapsto \a_{i,+} \psi_{+,i}, ~ \psi_{-,i} \mapsto \a_{i,-} \psi_{-,i}.$$
In order for this symmetry to preserve $\xoverline{Q}_+$, from \eq{eq:supercharge} we see that a necessary and sufficient condition is
\be\label{asym_sym_condition}
\a_{i}=\a_{i,+}~, ~\a_{i,-} {\partial_iW} (\underline{\a}\cdot \underline{\varphi}) = {\partial_iW} (\underline{\varphi})
\ee
Note that this is precisely the condition that the Lagrangian of the theory \eq{Lagrangian} is invariant under $g$. 
Moreover, such a symmetry always preserves a right-moving ${\cal N}=2$ superconformal algebra which is given by \eq{multifields_SCA} and \eq{eq:algebra} with $\psi_{-,i}$ replaced by $\psi_{+,i}$.
However it does not preserve the left-moving ${\cal N}=2$ superconformal algebra that acts on the $\xoverline{Q}_+$-cohomology. While $T_-$ and $J_-$ are always preserved, generically the supercurrents $G_-$ and $\bar G_-$ are not left invariant by the above $g$. Note that this however does not hinder the definition of the twined elliptic genus. 
To derive the elliptic genus twined by such a symmetry, we simply repeat the free field calculation \eq{free_1}--\eq{eq:product}. 
The result is the following. The elliptic genus of a LG orbifold with a superpotential of $N$ superfields, $\Phi_i$, $i=1\ldots N$,  twined under a symmetry of the Lagrangian which rotates the fields as 
\be
g: \varphi_i \mapsto e^{2\p i \l_i} \varphi_i, ~\psi_{+,i} \mapsto e^{2\p i \l_i} \psi_{+,i}, ~ \psi_{-,i} \mapsto e^{2\p i \l'_i} \psi_{-,i} ,
\ee
 is given by 
\begin{equation}\label{eq:orbeg_asym}
Z_g(\tau, z) = \frac{1}{h} \sum_{a, b \in \ZZ/h}(-1)^{\hat{c}(a + b + ab)} e^{2 \pi i (\hat{c}/2)(a^2 \tau + 2 a z)}\prod_{i=1}^N Z_{k_i,\l_i,\l'_i}(\tau, z + a \tau + b)
 \end{equation}
 with 
 \be
 Z_{k,\l,\l'} = \frac{\theta_1(\t, {k+1\over k+2}z- \l')}{\theta_1(\t,   {z\over k+2}+\l)}.\ee
 In particular, note that the condition that the boson $\varphi_i$ and the right-moving fermion $\psi_{+,i}$ have to be rotated by the same phase guarantees the holomorphicity of the twining genus, consistent with the interpretation of this condition as arising from the invariance of the supercharge $\xoverline{Q}_+$.

\begin{table}[htb]
\begin{center}
\begin{tabular}{C|C|C|C|C|C}
\text{Order of }g& \Pi_g & \textbf{EG}^c_g& X& \underline{\l}& \underline{\l'}\\\toprule
\multirow{2}*{2}& \multirow{2}*{$2^{12}$}&  \multirow{2}*{$\phi_{2B}^X$}& \multirow{2}*{$24A_1$}&1^22^2&1^6 \\
&&&&2^4/1^2 & 1^2 2^2\\\midrule
\multirow{2}*{4}& \multirow{2}*{$2^4 4^4$}& \multirow{2}*{$\phi_{4A}^X$}&\multirow{2}*{$24A_1$}&4^2/1^2&2^4/1^2\\
&&&&1^24^2/2^2 &1^22^2\\\midrule
6&6^4& \phi_{6B}^X&24A_1&2^1 6^1/1^2 &2^2 3^1/1^1\\\midrule
10& 2^2 10^2&\phi_{10A}^X&24A_1&2^2 5^1/1^3 &1^1 5^1\\\midrule
6&2^3 6^3&\phi^X_{6AD}&12A_2&1^12^16^1/3^1&1^5 6^1/2^13^1\\\midrule
4&4^6&\phi^X_{4B}&8A_3& \{0,0,0,{1\over 2},{1\over 4},{1\over 4}\}&1^22^2\\\midrule
4&4^2 8^2&\phi^X_{8A}&8A_3&2^28^1/1^24^1&2^14^1\\\midrule
14&1^12^17^114^1 &\phi^X_{14AB} &8A_3&  \{0,0,{1\over 2},{3\over 14},{5\over 14},{13\over 14}\}  &\{0,0,0,{1\over 7},{2\over 7},{4\over 7}\}\\
\end{tabular}\caption{\small Twining genera arising as in \eq{aym_twining_cubic}. We use the same notation as in Tables \ref{tbl:twinings} and \ref{tbl:Conway}. Whenever possible we encode the set of phases $\underline{\l}$ and $\underline{\l'}$ by the corresponding 6-dimensional Frame shapes. 
}\label{tbl:asymm1}
\end{center}
\end{table}

More explicitly, the above equation applied to a cubic superpotential leads to
\be\label{aym_twining_cubic}
\textbf{EG}^{c}_{g}(\tau, z)={1\over 3}\sum_{a,b\in \ZZ/3} q^{a^2}y^{2a}\prod_{i=1}^6 \frac{\theta_1\left (\t, {2\over 3}(z+a\t+b)-\l_i'\right)}{\theta_1\left (\t,{1\over 3}(z+a\t+b)+\l_i\right)}. 
\ee
In Table \ref{tbl:asymm1} we list some umbral moonshine twining functions that can arise in the above way through \eq{aym_twining_cubic}. 
Without going into the details of all of them, in what follows we describe how some of these phases arise as asymmetric symmetries of certain superpotentials. For instance, the phases in the first row of Table \ref{tbl:asymm1} can arise from keeping invariant the superfields $\Phi_1, \dots , \Phi_4$ while  rotating the component fields as
\be
\varphi_i \mapsto -\varphi_i, ~\psi_{+,i} \mapsto - \psi_{+,i}, ~ \psi_{-,i} \mapsto  \psi_{-,i}
\ee
for $i=5,6$. It is easy to check that such a transformation preserves the Lagrangian of the LG model with superpotential ${\cal W}^c_6$ (cf. Table \ref{tbl:groups_superpotentials}). Similarly, the phases in the second row of Table \ref{tbl:asymm1} can arise from transforming $\Phi_5$ and $\Phi_6$ as above, and simultaneously permuting $\Phi_1, \dots \Phi_4$ in two pairs of two. 
As a final example, we observe that the order 14 phases in Table \ref{tbl:asymm1} also arise as an asymmetric symmetry of the Fermat cubic theory by acting on the fields as 
\begin{align}\notag
&\underline{\varphi} \mapsto \text{diag}(1,1,-1,-\zeta_7^3,-\zeta_7^5, -\zeta_7^6 )\underline{\varphi}\notag
\\ &\underline{\psi_+} \mapsto \text{diag}(1,1,-1,-\zeta_7^3,-\zeta_7^5, -\zeta_7^6 )\underline{\psi_+}\notag
\\& \underline{\psi_-} \mapsto \text{diag}(1,1,1,\zeta_7,\zeta_7^4, \zeta_7^2 )\underline{\psi_+}\;.
\end{align}

\begin{table}
\begin{center}
\begin{tabular}{c|c|c|c|c|c}
Order of $g$ & $\Pi_{g}$  & $\textbf{EG}^q_g$&$X$& $\underline{\l}$  & $\underline{\l'}$\\\hline
\multirow{2}*{4}&\multirow{2}*{$4^6$}&\multirow{2}*{$\phi^X_{4C}$}&\multirow{2}*{$24A_1$}&$1^24/2$&$1^4$\\
&&&&$2.4/1^2$&$2^4/1^4$\\
\end{tabular}\caption{\small Twining genera arising as in \eq{asym_twining_quartic}. We use the same notation as in Tables \ref{tbl:twinings} and \ref{tbl:Conway}. Whenever possible we encode the set of phases $\underline{\l}$ and $\underline{\l'}$ by the corresponding 6-dimensional Frame shapes.  }\label{tbl:asym_quartic}
\end{center}
\end{table}

\begin{table}
\begin{center}
\begin{tabular}{c|c|c|c|c|c|c|c}
Order of $g$&$\Pi_g$  & $\textbf{EG}^{}_g$&$X$& $\underline{\l_1} $ &$\underline{\l_2}$  &$ \underline{\l_1'} $ &$\underline{\l_2'} $      \\\hline
3&$3^8$&$\phi^X_{3B}$&$24A_1$&$1^2$&$3/1$&$1^2$&$1^2$\\
4&$4^6$&$\phi^X_{4C}$&$24A_1$&$4/2$&$2$&$2^2/1^2$&$2$\\
\end{tabular}\caption{\small Twining genera arising as in \eq{eq:orbeg_asym} for models with $k_1=k_2=1$, $k_3=k_4=2$, or degree 6 hypersurfaces in $W\mathbb{P}_{1,1,2,2}$. Here $\underline{\l_1} $, $\underline{\l_1'} $  denotes the group action on the two chiral superfields of  charge 1 and $\underline{\l_2} $, $\underline{\l_2'} $ denotes that on the two fields of charge 2 in the corresponding weighted projective space.
 }\label{tbl:asym_weighted_projective}
 \end{center}
\end{table}

Similarly, the general expression \eq{eq:orbeg_asym} applied to the case of a quartic superpotential leads to the expression 
\be\label{asym_twining_quartic}
\textbf{EG}^{q}_{g}(\tau, z)={1\over 4}\sum_{a,b\in \ZZ/4} q^{a^2}y^{2a}\prod_{i=1}^4 \frac{\theta_1\left (\t, {3\over 4}(z+a\t+b)-\l_i\right)}{\theta_1\left (\t,{1\over 4}(z+a\t+b)+\l_i'\right)},
\ee
and in Table \ref{tbl:asym_quartic} we list some umbral moonshine twining functions that can arise in the above way through \eq{asym_twining_quartic}. Extending our analysis to  quasi-homogeneous cases corresponding to $K3$s in weighted projective space,  we obtain some umbral moonshine twining functions that can arise for  models corresponding to  degree 6 hypersurfaces in $W\mathbb{P}_{1,1,2,2}$. They are listed in Table \ref{tbl:asym_weighted_projective}. 
Note that the phases in Tables \ref{tbl:asymm1}--\ref{tbl:asym_weighted_projective} are all such that the resulting twining function has no pole as a function of $z$.

Without going further into the detailed analysis of these asymmetric symmetries, we close this subsection with the following observations. 
First, the set of twining genera we find in this subsection  appear to have a different relation to the Conway module (cf. \cite{DuncanMackCrane}) than the twining functions we obtained in \S \ref{sec:data}. Namely, not all of the twining genera arising in Tables \ref{tbl:asymm1}--\ref{tbl:asym_weighted_projective} find a counterpart among the proposed twined elliptic genera arising 
from the Conway module, though all of them coincide with certain twining genera arising from umbral moonshine \cite{UMk3}. Specifically, the functions $\phi^X_{3B}$ and $\phi^X_{6B}$ for $X=24A_1$, corresponding to $M_{24}$ elements with cycle shape $3^8$ and $6^4$ respectively, do not arise from the Conway module in the way proposed in \cite{DuncanMackCrane}. 
Similarly, $\phi^X_{4C}$ for $X=24A_1$, corresponding to $M_{24}$ elements with cycle shape $4^6$, does coincide with a Conway module twining function but the latter is attached to a group element of order 8. Note that these are the only 4-plane preserving umbral moonshine twining genera that cannot be obtained from the Conway module. Moreover, to the best of our knowledge, the present context is the first time we have seen these twining functions appearing as the twining genera of an actual symmetry of (the UV version of) a $K3$ sigma model. 

Second, we note a clear difference between the modular property of the twining genera arising from symmetries  in \S \ref{sec:data} and from asymmetric symmetries. 
Recall that all $\phi^X_{g}$ arising from umbral moonshine have the transformation property:
\be
\phi^X_{g}\left({a\t+b\over c\t+d},{z\over c\t+d}\right) e^{-2\pi i {cmz^2\over c\t+d}} = \psi(\gamma) \phi^X_{g}(\t,z)
\ee
for all 
\be
\gamma=\bem a&b \\c&d \eem \in \Gamma_0({\rm ord}(g))
\ee
and a certain group homomorphism $\psi: \Gamma_0({\rm ord}(g)) \to \CC^\ast$.
We say that $\psi$ is real if the image of $\psi$ lies in $\RR$, and say that $\psi$ is trivial if the image of  $\psi$ is $1$.
While all the twining functions arising from symmetries  in \S \ref{sec:data} have real multiplier systems, the twining genera arising from asymmetric twinings generally have complex multiplier systems.  
We will comment on the significance of such twining genera in the final section.

\subsection{Twinings of $M_{11}$}\label{sec:m11}

In \S\ref{sec:data} and \S\ref{sec:asymm} we focus on  symmetries arising from specific LG models. 
At the same time, recall that among all LG models with different superpotentials but the same number of chiral superfields with the same $U(1)$ charge, the (twined) elliptic genus always arises from the same free field expression given by \eq{eq:orbeg} and \eq{eq:orbegtwine}. 
This raises the following natural questions: is it possible to combine symmetries which are realized in different LG models with different superpotentials? Can we describe the twinings arising from a subgroup of a umbral group which is not a symmetry group of any explicit LG model in a uniform way? 
In this subsection we answer the question positively, by first noting that all twining functions $\phi_{g}^X$ arising from umbral moonshine for the case $X=12A_2$, where $[g]\subset 2.M_{12}$ has a representative in a copy of $M_{11}\subset 2.M_{12}$,  appear as the twining genus of {\em some} cubic LG model (cf. Table \ref{tbl:twinings}). In other words, all such functions $\phi_{g}^X$ admit an expression in terms of \eq{twining_cubic} with the choices of phases $\underline{\l}$ that are described in \S\ref{subsec:cubic} and Appendix \ref{app:groupactions}. Note that this is particularly interesting since no explicit cubic model has a symmetry group containing $M_{11}$. Hence this example might provide general hints about how symmetries realised in different points in the moduli space of $K3$ sigma models might be combined.  See also \cite{Taormina:2013jza} for an earlier discussion on combining (or surfing) symmetries of different Kummer surfaces.

Moreover, all the twined functions of $M_{11}\subset 2.M_{12}$ can be described uniformly in terms of a natural $12$-dimensional representation of the Mathieu group $M_{11}$. 
Explicitly, the relevant  $12$-dimensional representation is $\chi_1\oplus \chi_5$ in terms of irreducible representations of $M_{11}$
(cf. Table \ref{tbl:M11}). For convenience we also list in the character table the corresponding 12-dimensional Frame shape. 

Now, there is a unique way to split the 12 eigenvalues of a conjugacy class $[g]$ of $M_{11}$ into a set of 6 and their complex conjugate such that the following two conditions are satisfied\footnote{Clearly, not all subgroups of $S_{12}$ give rise to eigenvalues that can all be split in this way.}. Denote the 2 sets of phases, naturally defined mod $\ZZ$, by ${\L_g}$ and $\L_g^\ast$. The conditions are: 1) If an eigenvalue appears $k$ times in the 12-dimensional representation, then it must appear at least $\lfloor {k\over 2}\rfloor$ times in $\L_g$ as well as in  $\L_g^\ast$, 2) $\sum_{\l \in \L_g} \l \in {1\over 3} \ZZ$. 
The first condition says that $\L_g$ and $\L_g^\ast$ embodies the most symmetric possible split of the 12 eigenvalues, and the second ``torsion" condition should be related to the fact that our theory is a $\ZZ_3$-orbifold theory. In terms of these phases, if we define the twining function
\be\label{M11twine}
\f_{g}(\tau, z)={1\over 3 }\sum_{a,b\in \ZZ/3} q^{a^2}y^{2a} {2\over e^{2\pi i \sum_{\l \in \L_g} \l}+e^{-2\pi i \sum_{\l \in \L_g} \l}}
\prod_{\l\in \L_g}  \frac{\theta_1\left (\t, {2\over 3}(z+a\t+b)-\l\right)}{\theta_1\left (\t,{1\over 3}(z+a\t+b)+\l\right)}, 
\ee
then $\f_g=\f_{g'}^{12A_2}$ where the conjugacy $[g']$ is determined by the embedding of the group $M_{11}\subset 2.M_{12}$. 

Note that what we choose to call ${\L_g}$ and their conjugate  $\L_g^\ast$ is completely immaterial as long as the above-mentioned  two conditions are satisfied. 

As we will discuss in \S\ref{sec:con}, a description very similar to the above also applies for the subgroup $L_2(7)$ of $G(8A_3) \cong 2.AGL_3(2)$.

\section{Conclusions and Discussion}\label{sec:con}

First, we will start by discussing and summarizing the main results of this paper. 
\begin{enumerate}
\item
 One of our main motivations was to gather more data about the symmetries of  $K3$ sigma models and their corresponding twinings by using a Landau-Ginzburg orbifold description of  UV theories that flow to $K3$ sigma models, exploiting the fact that symmetries and twining genera are invariant under  RG flow. We find this approach valuable since we have very little computational control at generic points in the moduli space of  $K3$ sigma models; in fact, torus orbifolds furnish one of the few types of solvable models, but it is known that at these points in moduli space, the symmetries are far from generic \cite{Gaberdiel:2013nya}.  
Our investigation shows that this is indeed a rewarding approach. For instance, in \S \ref{sec:data} we presented explicit LG models with symmetries of order 11, 14 and 15 which preserve the full ${\cal N}=(4,4)$ superconformal symmetry in the IR. 
This is to the best of our knowledge the first time where these symmetries, though predicted to be realized at some isolated points in the moduli space according to lattice computations \cite{GHV}, have been found in explicit models. Moreover, we have explicitly computed their twinings and found them to coincide with the prediction of umbral moonshine with $X=12A_2$, $8A_3$, and $6D_4$, indicating the relevance of more than just the $M_{24}$ case of umbral moonshine for understanding symmetries of $K3$ CFTs.
\item A second motivation was the following. 
In \cite{UMk3} a relation between all 23 cases of umbral moonshine (not just the $X=24A_1$ case corresponding to Mathieu moonshine) and symmetries and twining genera of $K3$ sigma models, has been proposed. Although some consistency checks have been presented in \cite{UMk3}, one can certainly hope for more evidence for the existence of such a relation. From the data we collect in \S \ref{sec:data}-\ref{sec:moretwining} it is apparent that many of the umbral moonshine functions not arising from the $X=24A_1$ have been realized as twining genera of explicit symmetries of concrete models of $K3$ CFT. This lends support to the relation proposed in  \cite{UMk3}. 
Similarly, the fact that all the twining functions described in \S \ref{sec:data}, arising from symmetries that transform all components of the superfields in identical ways, can (also) be realized in the Conway module lends support to the relation between the Conway module and the $K3$ sigma models proposed in \cite{DuncanMackCrane}. 
Note that a vast majority, though not all, of the twining genera corresponding to symmetries preserving at least a 4-plane that are predicted by umbral moonshine can also be realized in the Conway module. This is a remarkable fact that we hope to understand better in the future.

\item In \S\ref{sec:asymm} we extended our analysis to include more general symmetries that preserve the $\xoverline{Q}_+$-cohomology (including its bi-grading), despite acting on different components of the chiral superfields in different ways. This property makes it possible to define the corresponding twining genera. 
By including these more general types of symmetries we recover many more twining genera predicted by umbral moonshine, including all those which do not arise from the Conway module. In particular, among the umbral moonshine twinings in Table \ref{tbl:asymm1}--\ref{tbl:asym_weighted_projective}, those corresponding to the Frame shapes $3^8, 6^4, 4^6, 4^28^2, 2^36^3, 2^210^2$ have not been found before in the context of $K3$ sigma models  as far as we know. 

Another interesting feature of the asymmetric symmetries is that only this type of symmetry can lead to twining genera with complex multiplier systems\footnote{We thank Roberto Volpato for a discussion on closely related matters.}, including all those mentioned above except for those corresponding to $2^36^3$ and  $2^210^2$.  In fact, to the best of our knowledge, the present context is the first time we have seen umbral twining functions with complex multiplier systems appearing as the twining genera of an actual symmetry of (the UV version of) a $K3$ sigma model. 
We know from general CFT arguments that the symmetries leading to twining functions with complex multiplier systems have to act on the theory in a rather intricate way. For instance, such a symmetry cannot be used to orbifold the theory since the resulting would-be twisted sectors would not satisfy the level-matching condition.  It is hence perhaps not surprising that we see such functions arising from the rather subtle asymmetric symmetries. 

The above observations suggest that a deeper understanding of these asymmetric symmetries may be crucial in unravelling the relation between $K3$ string theory and umbral moonshine.

\item 
The fact that all the different LG orbifold models with the same number of chiral superfields with the same  $U(1)$ charges have the same free field expression for their elliptic genus, irrespective of their superpotentials, suggests the possibility of combining symmetries realised at different points in the moduli space. In \eq{M11twine} we find that twining genera corresponding to all elements of $M_{11}\subset 2.M_{12}$, as dictated by the umbral moonshine with $X=12A_2$, can indeed be expressed in a uniform way in terms of a  natural 12-dimensional representation of $M_{11}$. 

 In fact, we can similarly consider $L_2(7) \subset 2.AGL_3(2)$ for the case $X=8A_3$. Note that $L_2(7)$ is the subgroup fixing one point in the 8-dimensional permutation representation of $AGL_3(2)$, and in this sense it is the exact counterpart of $M_{11}$.  The twining genera corresponding to all elements of $L_2(7)$, as dictated by  umbral moonshine with $X=8A_3$ and coinciding with what  we obtained from the quartic models in \S\ref{subsec:Quartics}, admit a uniform expression in terms of  a  natural 8-dimensional representation of $L_2(7)$.
 
 Clearly, it would be desirable to obtain a more directly physical interpretation of  \eq{M11twine}. 
Moreover, it would be very attractive if one could extend such a uniform description to the full umbral group ($2.M_{12}$ resp. $2.AGL_3(2)$ in the above cases), and to understand the geometric and/or physical meaning of such a uniform expression. 
\end{enumerate}

Finally we  close the main part of this paper with some comments on  open questions and possible future directions. 
\begin{enumerate}

\item
We  note that the asymmetric symmetries discussed in this paper -- symmetries which act differently on the bosonic and fermionic components of the chiral multiplets -- are reminiscent of the types of symmetries studied in \cite{KHP}. In that paper, the authors studied symmetries which arise in UV theories which flow in the IR to ${\cal N}=(0,4)$ superconformal theories with K3 target. These lie in the moduli space of worldsheet theories of  $E_8\times E_8$ heterotic string compactifications on K3. This moduli space is much more complicated than that of $(4,4)$ K3 sigma models; in fact, its global form is not known. This is because these theories involve a choice of embedding of 24 instantons into the $E_8\times E_8$ gauge group, arising from the requirement that the spacetime Bianchi identity for the three-form field strength $H$ of the heterotic string be satisfied. This involves a choice of stable, holomorphic vector bundles in the two $E_8$s, and the left-moving fermions couple to the gauge connections on these bundles.

One can construct such theories using $(0,2)$ UV gauged linear sigma models, along the lines of \cite{DK}, where the basic components are $(0,2)$ chiral and Fermi multiplets. When decomposing a $(2,2)$ chiral multiplet into $(0,2)$ multiplets, one finds that it decomposes as  a $(0,2)$ chiral multiplet which contains the boson and right-moving fermion of the $(2,2)$ chiral multiplet, and a $(0,2)$ Fermi multiplet, which contains the left-moving fermion of the $(2,2)$ chiral multiplet. From this point of view, the form of our asymmetric symmetries is highly suggestive of symmetries acting differently on the chiral and Fermi multiplets, which may naturally arise when deforming the bundle away from the standard embedding.  It would be interesting to explore this connection further.

\item 
A question that naturally arises from \cite{UMk3}  and the present work is the following. 
If the 23 cases of umbral moonshine, as well as the Conway module as proposed in \cite{DuncanMackCrane}, are all indeed relevant for the description of $K3$ sigma model symmetries and the corresponding twinings, how do we know when each case of umbral moonshine is relevant for describing a $K3$ CFT at a given point in moduli space? 

The classification theorem \cite{GHV}  (see also \cite{HuybrechtsDerived}) 
 of  symmetries preserving the ${\cal N}=(4,4)$ superconformal algebra of non-singular $K3$ sigma models  can be extended to include singular CFTs in the moduli space of $K3$ sigma models \cite{Cheng:2016org}. Using this one can prove that the symmetries of such a theory can always be embedded in one of the 23 umbral groups or in the Conway group.
This is in turn a consequence of the result that the $20$-dimensional lattice orthogonal in $H^\ast(K3,\ZZ) \cong \G^{4,20}$ to the four-dimensional subspace of signature $(4,0)$ determined by the CFT data can always be primitively embedded in one of the 23 Niemeier lattices or in the Leech lattice  \cite{Cheng:2016org}. Moreover it is plausible that for any one of the 23 Niemeier lattices $N(X)$, there exists at least a point in the CFT moduli space, necessarily corresponding to a singular CFT ${\cal T}$,  whose corresponding $20$-dimensional lattice can only be primitively embedded in $N$ \cite{Cheng:2016org}.  In this case it is natural to suspect that all the twining genera arising from symmetries of ${\cal T}$ can be captured by the umbral moonshine function corresponding to the Niemeier lattice $N(X)$. Through the compatibility of the lattice embeddings (and in particular the compatibility of the embedding of the root systems), the above conjecture connects the singularity type of the CFT to the Niemeier root systems determining the relevant cases of umbral moonshine for the particular CFT \cite{Cheng:2016org}.

For instance, we have seen that the umbral moonshine cases corresponding to the root systems $12A_2$ and $6D_4$ appear to be relevant for some of the cubic models we studied in \S\ref{subsec:cubic}, while the $8A_3$ case appears to be relevant for some of the quartic models we studied in \S\ref{subsec:Quartics}. 
Indeed, it is interesting to note a connection to the Landau-Ginzburg description of ADE-type minimal models $A_2, D_4$ in the case of cubic superpotentials, and $A_3$ in the case of quartic superpotentials \cite{Witten}. The LG description of these models has the UV superpotentials given by
\bea\nonumber
\mathcal W_{A_2} &=& X^3\\\nonumber
\mathcal W_{A_3} &=& X^4\\\nonumber
\mathcal W_{D_4} &=& X^3 + X Y^2.
\eea
Clearly, there appears to be a close connection between this singularity type from the point of view of the LG-- MM correspondence, and the singularity type arising from the root systems of the Niemeier lattices. It would be interesting to make this connection more precise.

\item While our results provide evidence for the  relation between $K3$ sigma models and umbral moonshine proposed in \cite{UMk3}, we know that umbral moonshine cannot be fully explained by only considering symmetries of $K3$ sigma models preserving  ${\cal N}=(4,4)$ superconformal symmetries. Clearly, such symmetries  necessarily correspond to subgroups of umbral groups which preserve at least a four-dimensional subspace in the natural 24-dimensional representation of the umbral group. 
Obviously, exploring the physical contexts in which the full umbral groups can arise without the 4-plane preserving constraint, will be an important necessary step for the physical understanding of umbral moonshine. 

\end{enumerate}

\section*{Acknowledgements}

We are indebted to John Duncan, Matthias Gaberdiel, Jeff Harvey, Gerald H\"ohn, Dan Isra\"el, Shamit Kachru and in particular Roberto Volpato for many useful discussions. 
MC is supported by ERC starting grant H2020 ERC StG 2014. 
SMH is supported by a Harvard University Golub Fellowship in the physical sciences. NMP is supported by a National Science Foundation Graduate Fellowship,
and also gratefully acknowledges the University of Amsterdam for hospitality
and the Delta Institute for Theoretical Physics for additional support while
this work was being completed. We would also like to thank the Perimeter Institute, Durham University, and Cambridge University for hospitality during the development of part of this work.

\appendix

\section{Connections to $K3^{[2]}$}\label{sec:K32}

	In this appendix, we will review some interesting results in algebraic geometry that explain the connection between symmetries of our cubic LG models and symplectic automorphisms of manifolds of type $K3^{[2]}$. If we attempt to interpret our cubic superpotentials geometrically via the usual CY/LG correspondence dictionary we encounter a puzzle: viewing the chiral superfields as coordinates in projective space, such a superpotential describes a hypersurface in $\mathbb{P}^5$, but the codimension is appropriate to describe a cubic fourfold in $\mathbb{P}^5$, not a $K3$ manifold. The upshot of our discussion will be the following: a large class of manifolds that are deformation equivalent to $K3^{[2]}$ can be described as the so-called Fano scheme of lines of a cubic fourfold. Symmetries of the Fano scheme are inherited from symmetries of the cubic equation describing the fourfold; in particular, symplectic automorphisms of the $K3^{[2]}$ correspond to supersymmetry-preserving symmetries of the CFTs that describe the IR fixed points of the cubic LG orbifold models. It is this connection that allows us to take advantage of the classification results in \cite{HohnMason}.

First, we define the Fano scheme of lines of a cubic fourfold. Let $X \subset\mathbb{P}^5$ be the cubic fourfold in $\mathbb{P}^5$. Then the Fano scheme (or variety) of lines is given by the following subvariety of the Grassmannian $Gr(\mathbb{P}^1, \mathbb{P}^5)$:
\begin{equation}\label{eq:Fano}
F(X)\equiv \left\lbrace [L] \in Gr(\mathbb{P}^1, \mathbb{P}^5) | L \in X \right\rbrace
\end{equation}
The automorphisms of the Fano scheme of lines of a fourfold descend from automorphisms of the fourfold itself, which can be viewed as a subgroup of $PGL(6, \mathbb{C})$ since the cubic is in 6 variables. Since we wish to find symplectic automorphisms, we will actually restrict to subgroups of $SL(6, \mathbb{C})$. It is a theorem of Beauville and Donagi \cite{BeauvilleDonagi} that the Fano scheme of lines is a simply connected 4(complex)-dimensional variety, with $H^{(2, 0)}(F(X)) = \mathbb{C}.\omega$, with $\omega$ a nowhere vanishing holomorphic 2-form. Moreover, there is an isomorphism of Hodge structures $H^4(X, \mathbb{Z}) \rightarrow H^2(F(X), \mathbb{Z})$. Fu \cite{Fu} classified automorphisms of the Fano scheme that are of primary order (order of the form $p^n$ where $p$ is prime) and that preserve the Pl{\"u}cker polarization (the natural polarization of $Gr(\mathbb{P}^1, \mathbb{P}^5)$) and it follows that such automorphisms always come from  automorphisms of the parent  cubic. More specifically, to find symplectic automorphisms of $F(X)$ one must restrict to  automorphisms of $X$ satisfying a certain condition. For us, this condition is the obvious one: $\sum_{j=1}^{6}\lambda_j \equiv 0 \textrm{ mod }\ZZ$, {\it i.e.} the product of all the six eigenvalues is one (cf. \eq{prod_is1}). See \cite{Fu, BeauvilleDonagi} for a rigorous, Hodge theoretic derivation of this condition.

While the approach of \cite{Fu} is useful for finding symmetries of certain prime orders, we would also like to understand the full group structure corresponding to particularly symmetric superpotentials. For this, we would like to use the classification of \cite{HohnMason} (see also \cite{Mongardi}) and therefore we must relate Fano schemes of lines in cubic fourfolds to our LG orbifolds.\footnote{For the relationship between cubic fourfolds and $K3$ sigma models in the more modern language of derived categories, see \cite{Huybrechts}.}

Happily, this relationship was explained in \cite{Cecotti} for the case of the Fermat superpotential, whose discussion we will briefly summarize, and which generalizes readily to general cubic superpotentials in six variables. There are 20 complex structure deformations of the Fermat superpotential of the form $a_{ijk} \Phi_{i} \Phi_{j} \Phi_{k}$ for $i \neq j \neq k$. (Recall that an algebraic $K3$ surface only has 19 complex structure deformations.) This cubic hypersurface has a complex structure moduli space of the form
\begin{equation}
{SO(2, 20) \over SO(2)\otimes SO(20)}.
\end{equation}
One can show that this moduli space is the same as the subspace of the $K3$ moduli space that is spanned by the chiral operators. Equivalently, in the language of \cite{Cecotti}, it is a certain moduli space of ``abstract" or ``superconformal" Hodge structures on $K3$, which incorporates the 20th complex structure deformation that is invisible in a polynomial describing an ordinary algebraic $K3$ surface. 

Moreover, if we have a so-called Pfaffian cubic, where the superpotential is given by the zero locus of a Pfaffian and only has 19 complex structure deformations, its Fano variety $F(X)$ will be isomorphic to $S^{[2]}$ for some $K3$ surface $S$. See \cite{Cecotti} for the details of the isomorphism between $F(X)$ and $S^{[2]}$. Here is the summary of the maps when $X$ is a Pfaffian cubic:

\begin{equation}
H^4(X) \xrightarrow{} H^2(F(X)) = H^2(S^{[2]}) \xrightarrow{\cong} H^2(S)\oplus \mathbb{C}.[E]
\end{equation}

On the right hand side, $[E]$ is the class of the exceptional divisor coming from the blow-up of the $A_1$ orbifold singularity on the $S^{[2]}$ moduli space; importantly, it is orthogonal to the forms inherited from the ``seed" $K3$ surface $S$, corresponding to the space $H^2(S)$, with respect to the canonical inner product. The same chain of isomorphisms turns out to hold in the case of more general cubic equations with 20 deformations. In particular, we can replace $S$ in the above maps by a general $K3$ sigma model and we can replace $S^{[2]}$ with a manifold that is deformation equivalent to a manifold of type $K3^{[2]}$.

To summarize: we expect that symplectic symmetries of the equation defining the cubic fourfold \cite{HohnMason} correspond to the symmetries of the UV LG superpotential given by the same equation. These symmetries act on the complex structure moduli of the UV theory, or equivalently on the Hodge structures of the $K3^{[2]}$ \cite{Cecotti}. By the chain of isomorphisms described above, we therefore expect these symmetries to persist in the IR, where they act on the chiral operators of the $K3$ SCFT.  Hence this explains why, from the above CFT and geometrical arguments, we expect the symmetry groups of \cite{HohnMason} to coincide with certain groups in the classification of \cite{GHV}.

Finally, in view of this connection it is natural to wonder about the relation between the twining genera of (a manifold that is deformation equivalent to a manifold of type) $K3^{[2]}$ and that of a $K3$ sigma model with the corresponding symmetry. 
Consider a symmetry $g$ of a $K3$ sigma model ${\cal T}$ and the corresponding symmetry $g'$ in the  symmetric orbifold theory ${\cal T}^{[2]}$. 
A straightforward generalization of the DMVV--Borcherds formula \cite{DMVV} gives the elliptic genus of ${\cal T}^{[2]}$ twined by $g'$ in terms of the twining genera of ${\cal T}$ corresponding to the symmetry $g$ via a lifting procedure  (see \cite{ChengM24K3}). In what follows we will discuss the specific example of an order 11 symmetry. 
Recall that the umbral moonshine prediction gives rise to two order 11 twining genera, corresponding to the cases $X=24 A_1$ and $X = 12 A_2$. They moreover coincide with the two order 11 twining genera arising from Conway module, corresponding to the two choices of sign for the constant $D_g$ (see Appendix \ref{sec:conwaytwinings}). 
Performing the explicit computation for our model $\mathcal{W}_1^c(\ubar{\Phi})$ yielded the the function corresponding to $X = 12 A_2$. Due to the above argument relating symmetries of the cubic model and those of $K3^{[2]}$ one might reasonably expect that the corresponding twining genus of  the $K3^{[2]}$-sigma model should be given by the lift of the $X = 12 A_2$ function. On the other hand, the twined elliptic genera of $K3^{[2]}$ computed in \cite{HohnMason} was shown to coincide with the lift of $\f^X_g$, $X=24A_1$, $[g]=11AB$. This is however simultaneously compatible with our expectation, it turns out that the two order 11 distinct twining genera $\f^X_{g}$, with $X = 24 A_1$ and $X = 12 A_2$ lift to the same equivariant genus of $K3^{[2]}$.\footnote{We are grateful to Gerald H{\"o}hn for correspondence on this point.} One may be able to understand this observation by the fact that order 11 cyclic groups have multiple orbits in $O(\Gamma_{4, 20})$ but only a single orbit in $O(\Gamma_{5, 21})$. 
It would be interesting to better understand the relationship between symmetries (and twinings) of $K3$ sigma models and those of manifolds of type $K3^{[2]}$ and $K3^{[n]}$ more generally. See \cite{HuybrechtsDerived} for the precise subset of $K3$ sigma model symmetries that also correspond to symplectic automorphisms of $K3^{[n]}$.

\section{Symmetries of LG Superpotentials}\label{app:groupactions}
In this Appendix we describe in more detail the symmetry groups of various superpotentials discussed in \S\ref{sec:data}, as well as the twining data that are summarized in the main text in Tables \ref{tbl:twinings},\ref{tbl:Conway},\ref{tbl:quartictwinings}. We write down the generators of the groups in the appropriate basis, when available, and delineate the conjugacy classes of these groups in Tables \ref{tbl:cubicW1}--\ref{tbl:cubicW6} for the cubic models, and Tables \ref{tbl:quarticW1}--\ref{tbl:quarticW4} for the quartic models.

\subsection{Cubics}
In \S \ref{subsec:cubic} we studied the twining genera arising as symmetries of the six maximal symmetry groups acting on six cubic superpotentials in six variables (cf. Table 
\ref{tbl:groups_superpotentials}). These symmetries were thoroughly discussed in \cite{HohnMason} (see also \cite{Mongardi}) and here we mostly summarize their descriptions for convenience while occasionally adding additional data for ease in reproducing our results.

\begin{enumerate}

\item \textbf{$L_2(11)$} \\
This group has order $2^2\cdot3\cdot 5 \cdot 11$.  The action of this group on the polynomial $ \Phi_1^2 \Phi_5 + \Phi_2^2 \Phi_4 + \Phi_3^2 \Phi_2 + \Phi_4^2 \Phi_1 + \Phi_5^2 \Phi_3 $ was proved in \cite{Adler}. It is not so hard to see that there is no symplectic extension obtained by adding the $\Phi_0^3$ term. A generating set of this group in a $5$-dimensional representation is given in \cite{ATLAS}:

\begin{align*}
\a& =
\left(\begin{array}{rrrrrr}
 0 & 1 & 0 & 0 & 0  \\
 1 & 0 & 0 & 0 & 0 \\
 0 & 0 & 0 & 0 & 1 \\
 1 & -1 & C & 1 & -C \\
 0 & 0 & 1 & 0 & 0
\end{array}\right) \\
\b&=
\left(\begin{array}{rrrrrr}
 0 & 0 & 0 & 1 & 0  \\
 0 & 0 & 1 & 0 & 0 \\
 0 & -1 & -1 & 0 & 0 \\
 -c & 0 & 0 & -1 & -2 c- C \\
 1 & 0 & 0 & -1 & 1
\end{array}\right) 
\end{align*}
where we have defined  $c= \zeta_{11} + \zeta_{11}^3 + \zeta_{11}^4 + \zeta_{11}^5 + \zeta_{11}^9$ and $C= -1- c$. 
Note that the basis used in the above matrices does not  correspond to the variables $\{\Phi_0, \Phi_1, \dots \Phi_5\}$ in which we write down the superpotential. However, since the eigenvalues are all the data we need in order to compute the twining genera, we are free to use whatever basis that is the most convenient. Of course, an appropriate change of basis should give us the order 11 and order 5 symmetries of the superpotential, given by, for instance, $g_{11} = \textrm{diag}(1, \zeta_{11}, \zeta_{11}^3, \zeta_{11}^4, \zeta_{11}^5, \zeta_{11}^9)$ and $g_{5}: (0)(12345)$ in the basis corresponds to $\{\Phi_0, \Phi_1, \dots \Phi_5\}$. The twining genera for this group are tabulated in Table \ref{tbl:cubicW1}.

\item \textbf{$(3 \times A_5)\!:\!2$}\\  This group has order $2^3\cdot3^2\cdot5$. The action of this group on the polynomial ${\cal W}_2^c$ was noted in \cite{Mongardi}; here we will provide a slightly different (but closely related) description of the group action. 
 To this end, we will pass to a basis of the chiral superfields for which the superpotential ${\cal W}^c_2$ takes the form
\be 
x_0^3 + x_1^3 + x_2^3 + x_3^3 - (x_0 + x_1 + x_2 + x_3)^3 + x_4^3 + x_5^3
\ee 
To understand the change of basis, recall that the hypersurface $y_0^2y_1 \ldots + y_3^2 y_0=0$ is in fact equivalent to Clebsch's cubic surface \cite{Mongardi}, which is expressible as the hypersurface $x_0^3 + x_1^3 + x_2^3 + x_3^3 - (x_0 + x_1 + x_2 + x_3)^3=0$.

The action of this group is now quite easy to see. Consider the 5-dimensional set $(x_0, x_1, x_2, x_3,  -(x_0 + x_1 + x_2 + x_3 ))$. The group $S_5$ acts in its natural way on this 5-dimensional set and so we can  identify the field transformations given by $A_5$ as permutations on this set. Put another way, we can always fix a change-of-basis matrix $A$ that takes the standard 5-dimensional representation via a similarity transformation to a 1+4-dimensional representation, and we can pick out the 4-dimensional block diagonal matrix to act on the $(x_0, \ldots, x_3)$ subset. 
Explicitly, we have \begin{align*}
A=\left(
\begin{array}{ccccc}
 1 & 1 & 1 & 1 & 1 \\
 -1 & 1 & 0 & 0 & 0 \\
 -1 & 0 & 1 & 0 & 0 \\
 -1 & 0 & 0 & 1 & 0 \\
 -1 & 0 & 0 & 0 & 1 \\
\end{array}
\right) .
\end{align*}
Using this matrix, the generators of $A_5$ are given in the basis $\{x_0,x_1,x_2,x_3\}$ by
\begin{align*}
g_1 &= 
\left(
\begin{array}{llll}
 -1 & 0 & 0 & 0 \\
 -1 & 0 & 1 & 0 \\
 -1 & 1 & 0 & 0 \\
 -1 & 0 & 0 & 1 \\
\end{array}
\right),
& g_2 &= 
\left(
\begin{array}{rrrr}
 1 & -1 & 0 & 0 \\
 0 & -1 & 0 & 1 \\
 0 & -1 & 1 & 0 \\
 0 & -1 & 0 & 0 \\
\end{array}
\right)
\end{align*}
to which we may append two rows/columns corresponding to the identity action on $x_4, x_5$. 
Now we need to add two more generators acting on all six fields to generate the complete group $(3 \times A_5)\!:\!2$. These can be given, in the basis of $\{x_0,\dots,x_5\}$, by
\begin{align*}
g_3 &= 
\left(
\begin{array}{llllll}
 1 & 0 & 0 & 0 & 0 & 0 \\
 0 & 1 & 0 & 0 & 0 & 0 \\
 0 & 0 & 1 & 0 & 0 & 0 \\
 0 & 0 & 0 & 1 & 0 & 0 \\
 0 & 0 & 0 & 0 & \zeta_3 & 0 \\
  0 & 0 & 0 & 0 & 0 & \zeta_3^{-1} \\
\end{array}
\right),
& g_4 &= 
\left(
\begin{array}{rrrrrr}
 0 & 0 & 0 & 1 & 0 & 0 \\
 1 & 0 & 0 & 0 & 0 & 0 \\
 0 & 1 & 0 & 0 & 0 & 0 \\
 0 & 0 & 1 & 0 & 0 & 0 \\
 0 & 0 & 0 & 0 & 0 & 1 \\
  0 & 0 & 0 & 0 & 1 & 0 \\
\end{array}
\right).
\end{align*}
The twining genera for this group are tabulated in Table \ref{tbl:cubicW2}.

\item \textbf{$A_7$}\\
This group has order $2^3\cdot3^2\cdot5\cdot 7$.
The action of this group can be easily described analogously to  the $A_5$ case above. 
In particular, the two standard generators of $A_7$ in the $6$-dimensional representation are given by 
\begin{align*}
g_1 &= \left(
\begin{array}{llllll}
  -1 & 1 & 0 & 0 & 0 & 0 \\
  -1 & 0 & 0 & 0 & 0 & 0 \\
  -1 & 0 & 1 & 0 & 0 & 0 \\
  -1 & 0 & 0 & 1 & 0 & 0 \\
  -1 & 0 & 0 & 0 & 1 & 0 \\
  -1 & 0 & 0 & 0 & 0 & 1 \\
\end{array}
\right),
&
g_2 &= \left(
\begin{array}{rrrrrr}
 1 & 0 & 0 & 0 & 0 & 0 \\
 0 & 0 & 1 & 0 & 0 & 0 \\
 0 & 0 & 0 & 1 & 0 & 0 \\
 0 & 0 & 0 & 0 & 1 & 0 \\
 0 & 0 & 0 & 0 & 0 & 1 \\
 0 & 1 & 0 & 0 & 0 & 0 \\
\end{array}
\right).
\end{align*}
The twining genera for this group are tabulated in Table \ref{tbl:cubicW3}.

\item \textbf{$M_{10}$}\\ The group has order $2^4\cdot3^2\cdot5$. This case has been described quite explicitly in \cite{HohnMason} and we reproduce their description here for completeness. $M_{10}$ is isomorphic to a certain extension of $A_6$, which we will denote by $3.A_6\langle \beta \rangle$   following \cite{HohnMason}. (More precisely, the projection of $3.A_6\langle \beta \rangle$ in $PSL(6, \mathbb{C})$, where $\beta$ normalizes $3.A_6$, is isomorphic to $M_{10}$.) There are two generators of $3.A_6$ which we call $\gamma_1, \gamma_2$. Together with $\beta$, they can be represented as $6 \times 6$ matrices in the basis of $\{\Phi_0,\dots,\Phi_5\}$ that leave $\mathcal{W}_4^c$ invariant. The generators are: 
\begin{align}
\gamma_1& = 
\left(\begin{array}{llllll}
 1 & 0 & 0 & 0 & 0 & 0 \\
 0 & 0 & 1 & 0 & 0 & 0 \\
 0 & 1 & 0 & 0 & 0 & 0\\
 0 & 0 & 0 & 0 & 1 & 0 \\
 0 & 0 & 0 & 1 & 0 & 0 \\
 0 & 0 & 0 & 0 & 0 & 1
\end{array}\right),
& \gamma_2& = 
\left(\begin{array}{rrrrrr}
 0 & 1 & 0 & 0 & 0 & 0 \\
 0 & 0 & \z_3 & 0 & 0 & 0 \\
 0 & 0 & 0 & 1 & 0 & 0\\
 \z_3^2 & 0 & 0 & 0 &0 & 0 \\
 0 & 0 & 0 & 0 & 0 & 1 \\
 0 & 0 & 0 & 0 & 1 & 0
\end{array}\right)\\
\beta& = {1 \over \sqrt{6}}
\left(\begin{array}{rrrrrr}
 1 & \z_3 & \z_3^2 & \z_3 & 1 & \z_3 \\
 \z_3^2 & 1 & 1 & \z_3 & \z_3 & \z_3 \\
 \z_3 & 1 & \z_3^2 & \z_3 & \z_3^2 & \z_3^2\\
 \z_3^2 & \z_3^2 & \z_3 & \z_3 & 1 & \z_3^2 \\
 1 & \z_3^2 & 1 & \z_3 & \z_3^2 & 1 \\
 \z_3^2 & \z_3^2 & \z_3^2 & \z_3^2 & \z_3^2 & \z_3
\end{array}\right).
\end{align}
The twining genera for this group are tabulated in Table \ref{tbl:cubicW4}.

\item \textbf{$3^{1 + 4}\!:\!2.2^2$}\\ 
This group has order $2^3\cdot3^5$. 
In this case, we start with the following group action $H = (3^2.S_3 \times 3^2.S_3).2$, coming from obvious permutations and multiplication by cube roots of unity. In addition, the superpotential is invariant under the following transformation $\alpha$:
\begin{align*}
\alpha& = {1 \over \sqrt{3}}
\left(\begin{array}{rrrrrr}
 \z_3 & \z_3^2 & 1 & 0 & 0 & 0 \\
 1 & 1 & 1 & 0 & 0 & 0 \\
 \z_3^2 & \z_3 & 1 & 0 & 0 & 0\\
 0 & 0 & 0 & \z_3^2 & \z_3 & 1 \\
 0 & 0 & 0 & \z_3^2 & \z_3^2 & \z_3^2 \\
 0 & 0 & 0 & \z_3^2 & 1 & \z_3
\end{array}\right).
\end{align*}
Together, $H\cap SL(6, \mathbb{Z})$ and $\alpha$ generate the group $3^{1 + 4}:2.2^2$.
The twining genera for this group are tabulated in Table \ref{tbl:cubicW5}.

\item \textbf{$3^4\!:\!A_6$}\\ This group has order $2^3\cdot3^6\cdot5$. The group action in this case is rather straightforward and has been already discussed in detail  in \cite{GHV} in the Gepner picture.  It is a combination of phase rotations by cube roots of unity and even permutations of the fields. Explicitly, the $A_6$ symmetries are the manifest even permutation symmetries of the chiral fields. To those, one can add the generators of the group $3^5$. These are, for example, all 20 diagonal matrices that have three entries of value $1$ and three $e^{2\pi i/3}$ entries. Finally, to account for the orbifold one quotients $3^5.A_6$ by the diagonal matrix that is $e^{2\pi i/3}$ times the $6\times6$ identity matrix and obtain $3^4.A_6$.

From a sigma model perspective, we restrict to even permutations of the fields since those preserve the full $\mathcal{N}=(4,4)$ SCA. In \cite{HohnMason}, this is equivalent to the condition that the  group action on the Fano scheme of lines in the corresponding cubic fourfold is {symplectic}; see Appendix \ref{sec:K32}. The twining genera for this group are tabulated in Table \ref{tbl:cubicW6}.

\end{enumerate}

\subsection{Quartics}\label{app:Quartics}
\begin{enumerate}

\item \textbf{$L_2(7)\times 2$}\\
This group has order $2^4\cdot3\cdot 7$.
 The so-called Klein's quartic (which lacks the lone $x_3^4$ term) has long been known to have an $L_2(7)$ symmetry. We reproduce the generators given in \cite{KleinBook}, to which one should append an extra row and column acting trivially on $x_3$:
\begin{align*}
g_1&= \left(\begin{array}{ccc}
 \zeta_7^4 & 0 & 0 \\
 0 & \zeta_7^2 & 0\\
 0 & 0 & \zeta_7 
\end{array}\right),
&
g_2 &= \left(\begin{array}{ccc}
 0 & 1 & 0 \\
 0 & 0 & 1\\
 1 & 0 & 0
\end{array}\right),
&
g_3 &= {-1 \over \sqrt{-7}}\left(\begin{array}{ccc}
 \zeta_7 - \zeta_7^6 & \zeta_7^2 - \zeta_7^5 & \zeta_7^4 -\zeta_7^3 \\
 \zeta_7^2 - \zeta_7^5 & \zeta_7^4 -\zeta_7^3 & \zeta_7 - \zeta_7^6\\
 \zeta_7^4 -\zeta_7^3 & \zeta_7 - \zeta_7^6 & \zeta_7^2 - \zeta_7^5
\end{array}\right).
\end{align*}
In addition, we append the following matrix to give the $\ZZ_2$ extension:
\begin{align*}
\left(\begin{array}{cccc}
 1 & 0 & 0 & 0 \\
 0 & 1 & 0 & 0\\
 0 & 0 & 1 & 0  \\
 0 & 0 & 0 & -1
\end{array}\right). 
\end{align*}
The twining genera for this group are tabulated in Table \ref{tbl:quarticW1}.

\item \textbf{$M_{20}$}\\
This group has order $2^6\cdot3\cdot 5$. 
 This group is simply that found by Mukai \cite{Mukai} as the symplectic automorphisms of a certain $K3$ surface, with the obvious action of $M_{20}\simeq 2^4:A_5$ induced from permutations and phase rotations. 
 The twining genera for this group are tabulated in Table \ref{tbl:quarticW2}.

\item \textbf{$T_{192}$}\\ This group has order $2^6\cdot3$. It is also among those found by Mukai \cite{Mukai} as the symplectic automorphisms of a certain $K3$ surface. It is isomorphic to $(Q_8 \times Q_8)\!:\!S_3$, where $Q_8$ is the quaternion group. The generators of the two copies of $Q_8$ are
\begin{align*}
g_1&= \left(\begin{array}{cc}
 I & 0 \\
 0 & id
\end{array}\right),
&
g_2 &= \left(\begin{array}{cc}
J&0 \\
0&id 
\end{array}\right),
&
g_3 &= \left(\begin{array}{cc}
 id & 0 \\
 0 & I
\end{array}\right),
&
g_4 &= \left(\begin{array}{cc}
id&0 \\
0&J
\end{array}\right),
\end{align*}
where $id$ is the $2\times2$ identity matrix, $I =  \left(\begin{array}{cc}
0&1 \\
-1&0
\end{array}\right)$, and $J =  \left(\begin{array}{cc}
i&0 \\
0&-i
\end{array}\right)$.
The two additional generators required are \cite{SmithThesis}:
\begin{align*}
g_5&= \left(\begin{array}{cccc}
 0 & 0 & 1 & 0  \\
 0 & 0 & 0 & 1 \\
 1 & 0 & 0 & 0 \\
 0 & 1 & 0 & 0
\end{array}\right),
&
g_6&= {(1 + i) \over 2}\left(\begin{array}{cccc}
 \zeta_3 & \zeta_3 & 0 & 0  \\
 i\zeta_3 & -i\zeta_3 & 0 & 0 \\
 0 & 0 & -i\zeta_3^2 & -\zeta_3^2 \\
 0 & 0 & -i\zeta_3^2 & \zeta_3^2
\end{array}\right).
\end{align*} 
The twining genera for this group are tabulated in Table \ref{tbl:quarticW3}.

\item \textbf{$(2 \times 4^2) : S_4$}\\
This group has order $2^8\cdot3$. 
 This Fermat example was  discussed in detail in the Gepner picture in \cite{GHV}. There is an obvious permutation symmetry of the fields given by $S_4$, while phase rotations act as $2 \times 4^2$.\footnote{This comes about after dividing out by a factor of $\mathbb{Z}_4$ accounting for the orbifold action-- the same phenomenon, dividing out by a phase rotation symmetry after orbifolding, also occurs in the $(1)^6$ Gepner model.} The twining genera for this group are tabulated in Table \ref{tbl:quarticW4}.

\end{enumerate}

\clearpage
\subsection{Tables}\label{subsec: twining tables}

\begin{table}[h]
\begin{center}
\caption{\label{tbl:cubicW1}Character table and twining functions of $L_2(11)$, with $\textbf{EG}^c_g = \phi^{12 A_2}_{h}$.}
\smallskip
\begin{footnotesize}
\begin{tabular}{c|rrrrrrrrr}\toprule
$[g]$&1A&2A&3A&5A&5B&6A&11A&11B\\
$[g^2]$&1A& 1A& 3A& 5B& 5A& 3A& 11B& 11A\\
$[g^3]$&1A& 2A& 1A& 5B& 5A& 2A& 11A & 11B\\
$[g^5]$&1A & 2A& 3A& 1A & 1A & 6A& 11A& 11B\\
$[g^{11}]$&1A&2A&3A&5A&5B&6A&1A&1A\\
\midrule
${\chi}_{1}$&1 & 1 & 1 & 1 & 1 & 1 & 1 & 1 \\
${\chi}_{2}$&5 & 1 & -1 & 0 & 0 & 1 & $\frac{1}{2} \left(-1+i \sqrt{11}\right)$ & $\frac{1}{2} \left(-1-i \sqrt{11}\right)$ \\
${\chi}_{3}$&5 & 1 & -1 & 0 & 0 & 1 & $\frac{1}{2} \left(-1-i \sqrt{11}\right)$ & $\frac{1}{2} \left(-1+i \sqrt{11}\right)$ \\
${\chi}_{4}$& 10 & -2 & 1 & 0 & 0 & 1 & -1 & -1 \\
${\chi}_{5}$&10 & 2 & 1 & 0 & 0 & -1 & -1 & -1 \\
${\chi}_{6}$& 11 & -1 & -1 & 1 & 1 & -1 & 0 & 0 \\
${\chi}_{7}$& 12 & 0 & 0 & $\frac{1}{2} \left(-1+\sqrt{5}\right)$ & $\frac{1}{2} \left(-1-\sqrt{5}\right)$ & 0 & 1 & 1 \\
${\chi}_{8}$&12 & 0 & 0 & $\frac{1}{2} \left(-1-\sqrt{5}\right)$ & $\frac{1}{2} \left(-1+\sqrt{5}\right)$ & 0 & 1 & 1 \\
\midrule
$[h]$&$1A$&$2B$& $3A$ & $5A$& $5A$ & $6C$ & $11AB$ &$11AB$  \\
\bottomrule
\end{tabular}
\end{footnotesize}
\end{center}
\end{table}
\begin{table}
\begin{center}
%\vspace{5pt}
%\smallskip
\caption{\label{tbl:cubicW2}Character table and twining functions of $(3 \times A_5):2$, with $\textbf{EG}^c_g = \phi^{6 D_4}_{h}$.
$A=\frac{1}{2} \left(-1-i \sqrt{15}\right)$, $\bar A= \frac{1}{2} \left(-1+i \sqrt{15}\right)$ }
\smallskip
\begin{footnotesize}
\begin{tabular}{c|rrrrrrrrrrrr}\toprule
$[g]$&1A& 6A& 6B & 2A & 4A & 5A & 15A & 15B & 3A & 3B & 2B & 3C\\
\midrule
${\chi}_{1}$& 1 & 1 & 1 & 1 & 1 & 1 & 1 & 1 & 1 & 1 & 1 & 1 \\
${\chi}_{2}$& 1 & 1 & -1 & -1 & -1 & 1 & 1 & 1 & 1 & 1 & 1 & 1 \\
${\chi}_{3}$&  2 & -1 & 0 & 0 & 0 & 2 & -1 & -1 & 2 & -1 & 2 & -1 \\
${\chi}_{4}$&  4 & 0 & 1 & -2 & 0 & -1 & -1 & -1 & 1 & 1 & 0 & 4 \\
${\chi}_{5}$&  4 & 0 & -1 & 2 & 0 & -1 & -1 & -1 & 1 & 1 & 0 & 4 \\
${\chi}_{6}$& 5 & 1 & -1 & -1 & 1 & 0 & 0 & 0 & -1 & -1 & 1 & 5 \\
${\chi}_{7}$& 5 & 1 & 1 & 1 & -1 & 0 & 0 & 0 & -1 & -1 & 1 & 5 \\
${\chi}_{8}$&6 & -2 & 0 & 0 & 0 & 1 & 1 & 1 & 0 & 0 & -2 & 6 \\
${\chi}_{9}$& 6 & 1 & 0 & 0 & 0 & 1 & $A$ & $\bar A$ & 0 & 0 & -2 & -3 \\
${\chi}_{10}$&  6 & 1 & 0 & 0 & 0 & 1 & $\bar A$ & $A$ & 0 & 0 & -2 & -3 \\
${\chi}_{11}$&8 & 0 & 0 & 0 & 0 & -2 & 1 & 1 & 2 & -1 & 0 & -4 \\
${\chi}_{12}$& 10 & -1 & 0 & 0 & 0 & 0 & 0 & 0 & -2 & 1 & 2 & -5 \\
\midrule
$[h]$&$1A$&$6A/6B$& $6A/6B$ & $2A/2B$& $4A$ & $5A$ & $15AB$ &$15AB$& $3A/3B$& $3A/3B$ & $2A/2B$& $3A/3B$  \\
\bottomrule
\end{tabular}
\end{footnotesize}
\end{center}
\end{table}

\begin{table}
\begin{center}
\caption{Character table and twining functions of $A_{7}$, with $\textbf{EG}^c_g = \phi^{24 A_1}_{h}$.}
\smallskip
\begin{footnotesize}
\begin{tabular}{c|rrrrrrrrr}\toprule
$[g]$&1A&2A&3A&3B&4A&5A&6A&7A&7B\\
$[g^2]$&1A&1A&3A&3B&2A&5A&3A&7A&7B\\
$[g^3]$&1A&2A&1A&1A&4A&5A&2A&7B&7A\\
$[g^5]$&1A& 2A& 3A & 3B & 4A& 1A& 6A& 7B& 7A\\
$[g^{7}]$& 1A&2A& 3A & 3B & 4A & 5A& 6A& 1A& 1A\\
\midrule
${\chi}_{1}$& 1 & 1 & 1 & 1 & 1 & 1 & 1 & 1 & 1 \\
${\chi}_{2}$&  6 & 2 & 3 & 0 & 0 & 1 & -1 & -1 & -1 \\
${\chi}_{3}$&  10 & -2 & 1 & 1 & 0 & 0 & 1 & $\frac{1}{2} \left(-1+i \sqrt{7}\right)$ & $\frac{1}{2} \left(-1-i \sqrt{7}\right)$ \\
${\chi}_{4}$&  10 & -2 & 1 & 1 & 0 & 0 & 1 & $\frac{1}{2} \left(-1-i \sqrt{7}\right)$ & $\frac{1}{2} \left(-1+i \sqrt{7}\right)$ \\
${\chi}_{5}$& 14 & 2 & 2 & -1 & 0 & -1 & 2 & 0 & 0 \\
${\chi}_{6}$&  14 & 2 & -1 & 2 & 0 & -1 & -1 & 0 & 0 \\
${\chi}_{7}$& 15 & -1 & 3 & 0 & -1 & 0 & -1 & 1 & 1 \\
${\chi}_{8}$&  21 & 1 & -3 & 0 & -1 & 1 & 1 & 0 & 0 \\
${\chi}_{9}$& 35 & -1 & -1 & -1 & 1 & 0 & -1 & 0 & 0 \\
\midrule
$[h]$&$1A$&$2A$& $3A$ & $3A$& $4B$ & $5A$ & $6A$ &$7AB$ & $7AB$ \\
\bottomrule
\end{tabular}\label{tbl:cubicW3}
\end{footnotesize}
\end{center}
\end{table}

\begin{table}
\begin{center}
\caption{Character table and twining functions of $M_{10}$, with $\textbf{EG}^c_g = \phi^{12A_2}_{h}$.}
\smallskip
\begin{footnotesize}
\begin{tabular}{c|rrrrrrrr}\toprule
$[g]$&1A&4A&4B&2A&3A&8A&8B&5A\\
$[g^2]$&1A& 2A& 2A& 1A& 3A& 4B& 4B& 5A\\
$[g^3]$&1A& 4A& 4B& 2A& 1A& 8A& 8B & 5A\\
$[g^5]$&1A & 4A& 4B& 2A & 3A & 8B& 8A& 1A\\
$[g^{7}]$&1A & 4A& 4B& 2A & 3A & 8B& 8A& 5A\\
\midrule
${\chi}_{1}$& 1 & 1 & 1 & 1 & 1 & 1 & 1 & 1 \\
${\chi}_{2}$& 1 & -1 & 1 & 1 & 1 & -1 & -1 & 1 \\
${\chi}_{3}$& 9 & -1 & 1 & 1 & 0 & 1 & 1 & -1 \\
${\chi}_{4}$&  9 & 1 & 1 & 1 & 0 & -1 & -1 & -1 \\
${\chi}_{5}$& 10 & 0 & -2 & 2 & 1 & 0 & 0 & 0 \\
${\chi}_{6}$& 10 & 0 & 0 & -2 & 1 & $-i \sqrt{2}$ & $i \sqrt{2}$ & 0 \\
${\chi}_{7}$&10 & 0 & 0 & -2 & 1 & $i \sqrt{2}$ & $-i \sqrt{2}$ & 0 \\
${\chi}_{8}$& 16 & 0 & 0 & 0 & -2 & 0 & 0 & 1 \\
\midrule
$[h]$&$1A$&$4C$& $4C$ & $2B$& $3A$ & $8CD$ & $8CD$ &$5A$  \\
\bottomrule
\end{tabular}\label{tbl:cubicW4}
\end{footnotesize}
\end{center}
\end{table}

\begin{sidewaystable}[h]
\begin{center}
\caption{Character table and twining functions of $3^{1+4}:2.2^2$, with $\textbf{EG}^c_g = \phi^{ 24A_1}_{h}$ with unhatted $h$. See Appendix \ref{sec:conwaytwinings} for those with hatted $h$.}
\smallskip
\begin{footnotesize}
\begin{tabular}{c|rrrrrrrrrrrrrrrrrrr}\toprule
$[g]$&1A& 3A& 3B& 3C& 3D& 3E& 3F& 3G& 3H& 3I& 3J& 3K& 2A& 6A& 4A& 4B& 12A& 12B& 4C\\
$[g^2]$&1A& 3A& 3B& 3C& 3D& 3E& 3F& 3G& 3H& 3I& 3J& 3K& 1A& 3A& 2A& 2A& 6A& 6A& 2A\\
$[g^3]$& 1A& 1A& 1A& 1A& 1A& 1A& 1A& 1A& 1A& 1A& 1A& 1A& 2A& 2A& 4A& 4B& 4B&4B& 4C\\
$[g^5]$&1A& 3A& 3B& 3C& 3D& 3E& 3F& 3G& 3H& 3I& 3J& 3K& 2A& 6A& 4A& 4B& 12B& 12A& 4C\\
$[g^{7}]$& 1A& 3A& 3B& 3C& 3D& 3E& 3F& 3G& 3H& 3I& 3J& 3K& 2A& 6A& 4A& 4B& 12B& 12A& 4C\\
$[g^{11}]$&1A& 3A& 3B& 3C& 3D& 3E& 3F& 3G& 3H& 3I& 3J& 3K& 2A& 6A& 4A& 4B& 12A& 12B& 4C \\
\midrule
$\chi_1$&1 & 1 & 1 & 1 & 1 & 1 & 1 & 1 & 1 & 1 & 1 & 1 & 1 & 1 & 1 & 1 & 1 & 1 & 1 \\
$\chi_2$& 1 & 1 & 1 & 1 & 1 & 1 & 1 & 1 & 1 & 1 & 1 & 1 & 1 & 1 & -1 & -1 & -1 & -1 & 1 \\
$\chi_3$& 1 & 1 & 1 & 1 & 1 & 1 & 1 & 1 & 1 & 1 & 1 & 1 & 1 & 1 & -1 & 1 & 1 & 1 & -1 \\
$\chi_4$& 1 & 1 & 1 & 1 & 1 & 1 & 1 & 1 & 1 & 1 & 1 & 1 & 1 & 1 & 1 & -1 & -1 & -1 & -1 \\
$\chi_5$& 2 & 2 & 2 & 2 & 2 & 2 & 2 & 2 & 2 & 2 & 2 & 2 & -2 & -2 & 0 & 0 & 0 & 0 & 0 \\
$\chi_6$& 8 & 8 & 5 & 2 & 2 & -1 & -1 & 2 & -1 & -1 & -4 & -4 & 0 & 0 & 0 & 0 & 0 & 0 & 0 \\
$\chi_7$& 8 & 8 & -4 & 2 & 2 & -1 & -1 & 2 & -1 & -1 & -4 & 5 & 0 & 0 & 0 & 0 & 0 & 0 & 0 \\
$\chi_8$& 8 & 8 & -4 & 2 & 2 & -1 & -1 & 2 & -1 & -1 & 5 & -4 & 0 & 0 & 0 & 0 & 0 & 0 & 0 \\
$\chi_9$& 8 & 8 & -1 & -1 & -1 & 8 & -1 & -1 & -1 & -1 & -1 & -1 & 0 & 0 & 0 & 0 & 0 & 0 & 0 \\
$\chi_{10}$& 8 & 8 & 2 & 5 & -4 & -1 & -1 & -4 & -1 & -1 & 2 & 2 & 0 & 0 & 0 & 0 & 0 & 0 & 0 \\
$\chi_{11}$& 8 & 8 & 2 & -4 & -4 & -1 & -1 & 5 & -1 & -1 & 2 & 2 & 0 & 0 & 0 & 0 & 0 & 0 & 0 \\
$\chi_{12}$& 8 & 8 & 2 & -4 & 5 & -1 & -1 & -4 & -1 & -1 & 2 & 2 & 0 & 0 & 0 & 0 & 0 & 0 & 0 \\
$\chi_{13}$& 8 & 8 & -1 & -1 & -1 & -1 & 8 & -1 & -1 & -1 & -1 & -1 & 0 & 0 & 0 & 0 & 0 & 0 & 0 \\
$\chi_{14}$& 8 & 8 & -1 & -1 & -1 & -1 & -1 & -1 & -1 & 8 & -1 & -1 & 0 & 0 & 0 & 0 & 0 & 0 & 0 \\
$\chi_{15}$& 8 & 8 & -1 & -1 & -1 & -1 & -1 & -1 & 8 & -1 & -1 & -1 & 0 & 0 & 0 & 0 & 0 & 0 & 0 \\
$\chi_{16}$& 18 & -9 & 0 & 0 & 0 & 0 & 0 & 0 & 0 & 0 & 0 & 0 & 2 & -1 & 0 & -2 & 1 & 1 & 0 \\
$\chi_{17}$& 18 & -9 & 0 & 0 & 0 & 0 & 0 & 0 & 0 & 0 & 0 & 0 & 2 & -1 & 0 & 2 & -1 & -1 & 0 \\
$\chi_{18}$& 18 & -9 & 0 & 0 & 0 & 0 & 0 & 0 & 0 & 0 & 0 & 0 & -2 & 1 & 0 & 0 & $\sqrt{3}$ & $-\sqrt{3}$ & 0 \\
 $\chi_{19}$&18 & -9 & 0 & 0 & 0 & 0 & 0 & 0 & 0 & 0 & 0 & 0 & -2 & 1 & 0 & 0 & $-\sqrt{3}$ & $\sqrt{3}$ & 0 \\
\midrule
$[h]$&$1A$&$\widehat{3C}$& $3A$ & $3A$& $3A$ & $3A$ & $3A$ &$3A$ & $3A$&$3A$&$3A$&$3A$&$2A$&$\widehat{6I}$&$4B$&$4B$&$\widehat{12L}$&$\widehat{12L'}$&$4B$ \\
\bottomrule
\end{tabular}\label{tbl:cubicW5}
\end{footnotesize}
\end{center}
\end{sidewaystable}

\begin{sidewaystable}[h]
\begin{center}
\caption{Character table and twining functions of $3^4:A_6$, with $\textbf{EG}^c_g = \phi^{24A_1}_h$ with unhatted $h$. See Appendix \ref{sec:conwaytwinings} for those with hatted $h$. \\
$A=\frac{1}{2} \left(-1-3 i \sqrt{3}\right)$, $\xoverline{A}=\frac{1}{2} \left(-1+3 i \sqrt{3}\right)$;
$B =\frac{1}{2} \left(1-\sqrt{5}\right)$, $\xoverline{B}= \frac{1}{2}\left(1+\sqrt{5}\right)$}
\smallskip
\begin{footnotesize}
\begin{tabular}{c|rrrrrrrrrrrrrrrrrrrrr}\toprule
$[g]$&1A& 3A& 3B& 3C& 2A& 6A& 6B& 6C& 4A& 3D& 9A& 9B& 3E& 3F& 3G& 9C& 9D& 3H& 3I& 5A& 5B\\
$[g^2]$&1A& 3A& 3B& 3C& 1A& 3A& 3B& 3C& 2A& 3D& 9B& 9A& 3E& 3F& 3G& 9D& 9C& 3H& 3I& 5B& 5A \\
$[g^3]$& 1A& 1A& 1A& 1A& 2A& 2A& 2A& 2A& 4A& 1A& 3A& 3A& 1A& 1A& 1A& 3A& 3A&1A& 1A& 5B& 5A\\
$[g^5]$&1A& 3A& 3B& 3C& 2A& 6A& 6B& 6C& 4A& 3D& 9B& 9A& 3E&3F& 3G& 9D& 9C& 3H& 3I& 1A& 1A\\
$[g^{7}]$& 1A& 3A& 3B& 3C& 2A& 6A& 6B& 6C& 4A& 3D& 9A& 9B& 3E& 3F& 3G& 9C& 9D& 3H& 3I& 5B& 5A\\
\midrule
$\chi_1$&1 & 1 & 1 & 1 & 1 & 1 & 1 & 1 & 1 & 1 & 1 & 1 & 1 & 1 & 1 & 1 & 1 & 1 & 1 & 1 & 1 \\
$\chi_2$& 5 & 5 & 5 & 5 & 1 & 1 & 1 & 1 & -1 & -1 & -1 & -1 & -1 & -1 & 2 & 2 & 2 & 2 & 2 & 0 & 0 \\
 $\chi_3$&5 & 5 & 5 & 5 & 1 & 1 & 1 & 1 & -1 & 2 & 2 & 2 & 2 & 2 & -1 & -1 & -1 & -1 & -1 & 0 & 0 \\
$\chi_4$& 8 & 8 & 8 & 8 & 0 & 0 & 0 & 0 & 0 & -1 & -1 & -1 & -1 & -1 & -1 & -1 & -1 & -1 & -1 & $B$ & $\xoverline{B}$ \\
$\chi_5$& 8 & 8 & 8 & 8 & 0 & 0 & 0 & 0 & 0 & -1 & -1 & -1 & -1 & -1 & -1 & -1 & -1 & -1 & -1 & $\xoverline{B}$ & ${B}$ \\
$\chi_6$& 9 & 9 & 9 & 9 & 1 & 1 & 1 & 1 & 1 & 0 & 0 & 0 & 0 & 0 & 0 & 0 & 0 & 0 & 0 & -1 & -1 \\
 $\chi_7$& 10 & 10 & 10 & 10 & -2 & -2 & -2 & -2 & 0 & 1 & 1 & 1 & 1 & 1 & 1 & 1 & 1 & 1 & 1 & 0 & 0 \\
$\chi_8$& 20 & -7 & 2 & 2 & -4 & -1 & 2 & 2 & 0 & 2 & -1 & -1 & 2 & 2 & 2 & -1 & -1 & 2 & 2 & 0 & 0 \\
 $\chi_9$&20 & -7 & 2 & 2 & 4 & 1 & -2 & -2 & 0 & 2 & -1 & -1 & 2 & 2 & 2 & -1 & -1 & 2 & 2 & 0 & 0 \\
$\chi_{10}$& 30 & 3 & 3 & -6 & 2 & -1 & -1 & 2 & 0 & -3 & 0 & 0 & 6 & -3 & 0 & 0 & 0 & 0 & 0 & 0 & 0 \\
$\chi_{11}$& 30 & 3 & 3 & -6 & 2 & -1 & -1 & 2 & 0 & -3 & 0 & 0 & -3 & 6 & 0 & 0 & 0 & 0 & 0 & 0 & 0 \\
$\chi_{12}$& 30 & 3 & 3 & -6 & 2 & -1 & -1 & 2 & 0 & 6 & 0 & 0 & -3 & -3 & 0 & 0 & 0 & 0 & 0 & 0 & 0 \\
$\chi_{13}$& 30 & 3 & -6 & 3 & 2 & -1 & 2 & -1 & 0 & 0 & 0 & 0 & 0 & 0 & 6 & 0 & 0 & -3 & -3 & 0 & 0 \\
 $\chi_{14}$&30 & 3 & -6 & 3 & 2 & -1 & 2 & -1 & 0 & 0 & 0 & 0 & 0 & 0 & -3 & 0 & 0 & -3 & 6 & 0 & 0 \\
$\chi_{15}$& 30 & 3 & -6 & 3 & 2 & -1 & 2 & -1 & 0 & 0 & 0 & 0 & 0 & 0 & -3 & 0 & 0 & 6 & -3 & 0 & 0 \\
$\chi_{16}$& 40 & -14 & 4 & 4 & 0 & 0 & 0 & 0 & 0 & -2 & 1 & 1 & -2 & -2 & 1 & $A$ & $\xoverline{A}$ & 1
   & 1 & 0 & 0 \\
$\chi_{17}$& 40 & -14 & 4 & 4 & 0 & 0 & 0 & 0 & 0 & -2 & 1 & 1 & -2 & -2 & 1 & $\xoverline{A}$ & $A$ & 1
   & 1 & 0 & 0 \\
$\chi_{18}$& 40 & -14 & 4 & 4 & 0 & 0 & 0 & 0 & 0 & 1 & $A$ & $\xoverline{A}$ & 1 & 1 & -2 & 1 & 1 & -2
   & -2 & 0 & 0 \\
$\chi_{19}$& 40 & -14 & 4 & 4 & 0 & 0 & 0 & 0 & 0 & 1 & $\xoverline{A}$ & $A$ & 1 & 1 & -2 & 1 & 1 & -2
   & -2 & 0 & 0 \\
$\chi_{20}$& 90 & 9 & -18 & 9 & -2 & 1 & -2 & 1 & 0 & 0 & 0 & 0 & 0 & 0 & 0 & 0 & 0 & 0 & 0 & 0 & 0 \\
$\chi_{21}$& 90 & 9 & 9 & -18 & -2 & 1 & 1 & -2 & 0 & 0 & 0 & 0 & 0 & 0 & 0 & 0 & 0 & 0 & 0 & 0 & 0 \\
\midrule
$[h]$&$1A$&$\widehat{3C}$& $3A$ & $3A$& $2A$ & $\widehat{6I}$ & $6A$ &$6A$ & $4B$& $3A$& $\widehat{9C}$&$\widehat{9C}$&$3A$&$3A$&$3A$&$\widehat{9C'}$&$\widehat{9C'}$&$3A$&$3A$&$5A$&$5A$  \\
\bottomrule
\end{tabular}
\label{tbl:cubicW6}
\end{footnotesize}
\end{center}
\end{sidewaystable}

\begin{table}
\begin{center}
\caption{Character table and twining functions of $L_2(7)\times2$, with $\textbf{EG}^q_g = \phi^{ 8A_3}_{h}$.\\
$A =\frac{1}{2} \left(1-i \sqrt{7}\right)$, $\bar A=\frac{1}{2}\left(1+i \sqrt{7}\right)$ }
\smallskip
\begin{footnotesize}
\begin{tabular}{c|rrrrrrrrrrrr}\toprule
$[g]$&1A& 3A& 2A& 6A& 4A& 4B& 14A&  7A& 14B&  7B& 2B& 2C \\
\midrule
$\chi_1$& 1 & 1 & 1 & 1 & 1 & 1 & 1 & 1 & 1 & 1 & 1 & 1 \\
$\chi_2$& 1 & 1 & -1 & -1 & -1 & 1 & -1 & 1 & -1 & 1 & -1 & 1 \\
$\chi_3$& 3 & 0 & -3 & 0 & -1 & 1 & $A$ & $-A$ & $\bar A$ &
   $-\bar A$ & 1 & -1 \\
$\chi_4$& 3 & 0 & -3 & 0 & -1 & 1 & $\bar A$ & $-\bar A$ & $A$ &
   $-A$ & 1 & -1 \\
$\chi_5$& 3 & 0 & 3 & 0 & 1 & 1 & $-A$ & $-A$ & $-\bar A$ &
   $-\bar A$ & -1 & -1 \\
$\chi_6$& 3 & 0 & 3 & 0 & 1 & 1 & $-\bar A$ & $-\bar A$ & $-A$ &
   $-A$& -1 & -1 \\
$\chi_7$& 6 & 0 & 6 & 0 & 0 & 0 & -1 & -1 & -1 & -1 & 2 & 2 \\
$\chi_8$& 6 & 0 & -6 & 0 & 0 & 0 & 1 & -1 & 1 & -1 & -2 & 2 \\
$\chi_9$& 7 & 1 & 7 & 1 & -1 & -1 & 0 & 0 & 0 & 0 & -1 & -1 \\
$\chi_{10}$& 7 & 1 & -7 & -1 & 1 & -1 & 0 & 0 & 0 & 0 & 1 & -1 \\
$\chi_{11}$& 8 & -1 & 8 & -1 & 0 & 0 & 1 & 1 & 1 & 1 & 0 & 0 \\
$\chi_{12}$& 8 & -1 & -8 & 1 & 0 & 0 & -1 & 1 & -1 & 1 & 0 & 0 \\
\midrule
$[h]$&$1A$&$3A$& $2A/2C$ & $6A$& $4C$ & $4C$ & $14AB$ &$7AB$&$14AB$&$7AB$&$2A/2C$&$2A/2C$  \\
\bottomrule
\end{tabular}
\label{tbl:quarticW1}
\end{footnotesize}
\end{center}
\end{table}

\begin{table}
\begin{center}
\caption{Character table and twining functions of $M_{20}$, with $\textbf{EG}^q_g = \phi^{24A_1}_{h}$.}
\smallskip
\begin{footnotesize}
\begin{tabular}{c|rrrrrrrrr}\toprule
$[g]$&1A& 2A& 2B& 4A& 4B& 4C& 3A& 5A& 5B\\
$[g^2]$&1A& 1A& 1A& 2A& 2A& 2A& 3A& 5B& 5A\\
$[g^3]$&1A& 2A& 2B& 4A& 4B& 4C& 1A& 5B& 5A\\
$[g^5]$&1A& 2A& 2B& 4A& 4B& 4C& 3A& 1A& 1A\\
\midrule
$\chi_1$&1 & 1 & 1 & 1 & 1 & 1 & 1 & 1 & 1 \\
$\chi_2$& 3 & 3 & -1 & -1 & -1 & -1 & 0 & $\frac{1}{2} \left(1-\sqrt{5}\right)$ & $\frac{1}{2} \left(1+\sqrt{5}\right)$ \\
$\chi_3$& 3 & 3 & -1 & -1 & -1 & -1 & 0 & $\frac{1}{2} \left(1+\sqrt{5}\right)$ & $\frac{1}{2} \left(1-\sqrt{5}\right)$ \\
$\chi_4$& 4 & 4 & 0 & 0 & 0 & 0 & 1 & -1 & -1 \\
$\chi_5$& 5 & 5 & 1 & 1 & 1 & 1 & -1 & 0 & 0 \\
$\chi_6$& 15 & -1 & 3 & -1 & -1 & -1 & 0 & 0 & 0 \\
$\chi_7$&15 & -1 & -1 & 3 & -1 & -1 & 0 & 0 & 0 \\
$\chi_8$& 15 & -1 & -1 & -1 & 3 & -1 & 0 & 0 & 0 \\
$\chi_9$& 15 & -1 & -1 & -1 & -1 & 3 & 0 & 0 & 0 \\
\midrule
$[h]$&$1A$&$2A$& $2A$ & $4B$& $4B$ & $4B$ & $3A$ &$5A$&$5A$  \\
\bottomrule
\end{tabular}
\label{tbl:quarticW2}
\end{footnotesize}
\end{center}
\end{table}

\begin{table}
\begin{center}
\caption{Character table and twining functions of $T_{192}$, with $\textbf{EG}^q_g = \phi^{24A_1}_{h}$.}
\smallskip
\begin{footnotesize}
\begin{tabular}{c|rrrrrrrrrrrrr}\toprule
$[g]$&1A& 2A& 3A& 4A& 2B& 4B& 4C& 2C& 6A& 2D& 2E& 2F& 4D\\
$[g^2]$&1A& 1A& 3A& 2B& 1A& 2D& 2F& 1A& 3A& 1A& 1A& 1A& 2E\\
$[g^3]$&1A& 2A& 1A& 4A& 2B& 4B& 4C& 2C& 2B& 2D& 2E& 2F& 4D\\
$[g^5]$&1A& 2A& 3A& 4A& 2B& 4B& 4C& 2C& 6A& 2D& 2E& 2F& 4D\\
\midrule
$\chi_{1}$&1 & 1 & 1 & 1 & 1 & 1 & 1 & 1 & 1 & 1 & 1 & 1 & 1 \\
$\chi_{2}$& 1 & -1 & 1 & 1 & 1 & -1 & -1 & -1 & 1 & 1 & 1 & 1 & -1 \\
$\chi_{3}$& 2 & 0 & -1 & 2 & 2 & 0 & 0 & 0 & -1 & 2 & 2 & 2 & 0 \\
$\chi_{4}$& 3 & -1 & 0 & -1 & 3 & -1 & 1 & -1 & 0 & 3 & -1 & -1 & 1 \\
$\chi_{5}$& 3 & -1 & 0 & -1 & 3 & 1 & 1 & -1 & 0 & -1 & 3 & -1 & -1 \\
$\chi_{6}$& 3 & 1 & 0 & -1 & 3 & -1 & -1 & 1 & 0 & -1 & 3 & -1 & 1 \\
$\chi_{7}$& 3 & 1 & 0 & -1 & 3 & 1 & -1 & 1 & 0 & 3 & -1 & -1 & -1 \\
$\chi_{8}$& 3 & -1 & 0 & -1 & 3 & 1 & -1 & -1 & 0 & -1 & -1 & 3 & 1 \\
$\chi_{9}$& 3 & 1 & 0 & -1 & 3 & -1 & 1 & 1 & 0 & -1 & -1 & 3 & -1 \\
$\chi_{10}$& 4 & 2 & 1 & 0 & -4 & 0 & 0 & -2 & -1 & 0 & 0 & 0 & 0 \\
$\chi_{11}$& 4 & -2 & 1 & 0 & -4 & 0 & 0 & 2 & -1 & 0 & 0 & 0 & 0 \\
$\chi_{12}$& 6 & 0 & 0 & 2 & 6 & 0 & 0 & 0 & 0 & -2 & -2 & -2 & 0 \\
$\chi_{13}$& 8 & 0 & -1 & 0 & -8 & 0 & 0 & 0 & 1 & 0 & 0 & 0 & 0 \\
\midrule
$[h]$&$1A$&$2A$& $3A$ & $4B$& $2A$ & $4B$ & $4B$ &$2A$&$6A$&$2A$&$2A$&$2A$&$4B$  \\
\bottomrule
\end{tabular}
\label{tbl:quarticW3}
\end{footnotesize}
\end{center}
\end{table}

\newgeometry{left=3cm,bottom=0.1cm}
\begin{sidewaystable}
\begin{center}
\caption{Character table and twining functions of $(2\times 4^2): S_4$, with $\textbf{EG}^q_g = \phi^{24A_1}_{h}$ for unhatted $h$. See Appendix \ref{sec:conwaytwinings} for those with hatted $h$. \\}
\smallskip
\begin{footnotesize}
\begin{tabular}{c|rrrrrrrrrrrrrrrrrrrrrrrr}\toprule
$[g]$&1A& 2A& 2B& 4A& 4B& 4C& 2C& 4D& 2D& 4E& 2E& 4F& 3A& 6A& 2F& 2G& 4G& 4H& 8A& 8B& 4I& 4J& 4K& 4L\\
$[g^2]$&1A& 1A& 1A& 2B& 2B& 2B& 1A& 2B& 1A& 2B& 1A& 2B& 3A& 3A& 1A& 1A& 2B& 2B& 4B& 4B& 2C& 2C& 2E& 2E \\
$[g^3]$&1A& 2A& 2B& 4A& 4B& 4C& 2C& 4D& 2D& 4E& 2E& 4F& 1A& 2A& 2F& 2G& 4G& 4H& 8A& 8B& 4I& 4J& 4K& 4L \\
$[g^5]$&1A& 2A& 2B& 4A& 4B& 4C& 2C& 4D& 2D& 4E& 2E& 4F& 3A& 6A& 2F& 2G& 4G& 4H& 8A& 8B& 4I& 4J& 4K& 4L \\
$[g^7]$& 1A& 2A& 2B& 4A& 4B& 4C& 2C& 4D& 2D& 4E& 2E& 4F& 3A& 6A& 2F& 2G& 4G& 4H& 8A& 8B& 4I& 4J& 4K& 4L \\
\midrule
$\chi_{1}$&1 & 1 & 1 & 1 & 1 & 1 & 1 & 1 & 1 & 1 & 1 & 1 & 1 & 1 & 1 & 1 & 1 & 1 & 1 & 1 & 1 & 1 & 1 & 1 \\
$\chi_{2}$& 1 & -1 & 1 & -1 & 1 & -1 & 1 & -1 & -1 & 1 & 1 & -1 & 1 & -1 & -1 & 1 & 1 & -1 & 1 & -1 & -1 & 1 & 1 & -1 \\
$\chi_{3}$& 1 & -1 & 1 & -1 & 1 & -1 & 1 & -1 & -1 & 1 & 1 & -1 & 1 & -1 & 1 & -1 & -1 & 1 & -1 & 1 & 1 & -1 & -1 & 1 \\
$\chi_{4}$& 1 & 1 & 1 & 1 & 1 & 1 & 1 & 1 & 1 & 1 & 1 & 1 & 1 & 1 & -1 & -1 & -1 & -1 & -1 & -1 & -1 & -1 & -1 & -1 \\
$\chi_{5}$& 2 & 2 & 2 & 2 & 2 & 2 & 2 & 2 & 2 & 2 & 2 & 2 & -1 & -1 & 0 & 0 & 0 & 0 & 0 & 0 & 0 & 0 & 0 & 0 \\
$\chi_{6}$& 2 & -2 & 2 & -2 & 2 & -2 & 2 & -2 & -2 & 2 & 2 & -2 & -1 & 1 & 0 & 0 & 0 & 0 & 0 & 0 & 0 & 0 & 0 & 0 \\
$\chi_{7}$& 3 & -3 & 3 & -3 & 3 & -3 & -1 & 1 & 1 & -1 & -1 & 1 & 0 & 0 & -1 & 1 & 1 & -1 & 1 & -1 & 1 & -1 & -1 & 1 \\
$\chi_{8}$& 3 & -3 & 3 & -3 & 3 & -3 & -1 & 1 & 1 & -1 & -1 & 1 & 0 & 0 & 1 & -1 & -1 & 1 & -1 & 1 & -1 & 1 & 1 & -1 \\
$\chi_{9}$& 3 & -3 & 3 & 1 & -1 & 1 & 3 & -3 & 1 & -1 & -1 & 1 & 0 & 0 & -1 & 1 & 1 & -1 & -1 & 1 & -1 & 1 & -1 & 1 \\
$\chi_{10}$&3 & -3 & 3 & 1 & -1 & 1 & 3 & -3 & 1 & -1 & -1 & 1 & 0 & 0 & 1 & -1 & -1 & 1 & 1 & -1 & 1 & -1 & 1 & -1 \\
$\chi_{11}$& 3 & 3 & 3 & -1 & -1 & -1 & 3 & 3 & -1 & -1 & -1 & -1 & 0 & 0 & -1 & -1 & -1 & -1 & 1 & 1 & -1 & -1 & 1 & 1 \\
$\chi_{12}$& 3 & 3 & 3 & -1 & -1 & -1 & 3 & 3 & -1 & -1 & -1 & -1 & 0 & 0 & 1 & 1 & 1 & 1 & -1 & -1 & 1 & 1 & -1 & -1 \\
$\chi_{13}$& 3 & 3 & 3 & 3 & 3 & 3 & -1 & -1 & -1 & -1 & -1 & -1 & 0 & 0 & -1 & -1 & -1 & -1 & -1 & -1 & 1 & 1 & 1 & 1 \\
$\chi_{14}$& 3 & 3 & 3 & 3 & 3 & 3 & -1 & -1 & -1 & -1 & -1 & -1 & 0 & 0 & 1 & 1 & 1 & 1 & 1 & 1 & -1 & -1 & -1 & -1 \\
 $\chi_{15}$&3 & -3 & 3 & 1 & -1 & 1 & -1 & 1 & 1 & -1 & 3 & -3 & 0 & 0 & -1 & 1 & 1 & -1 & -1 & 1 & 1 & -1 & 1 & -1 \\
$\chi_{16}$& 3 & -3 & 3 & 1 & -1 & 1 & -1 & 1 & 1 & -1 & 3 & -3 & 0 & 0 & 1 & -1 & -1 & 1 & 1 & -1 & -1 & 1 & -1 & 1 \\
 $\chi_{17}$&3 & 3 & 3 & -1 & -1 & -1 & -1 & -1 & -1 & -1 & 3 & 3 & 0 & 0 & -1 & -1 & -1 & -1 & 1 & 1 & 1 & 1 & -1 & -1 \\
$\chi_{18}$& 3 & 3 & 3 & -1 & -1 & -1 & -1 & -1 & -1 & -1 & 3 & 3 & 0 & 0 & 1 & 1 & 1 & 1 & -1 & -1 & -1 & -1 & 1 & 1 \\
$\chi_{19}$& 6 & 6 & 6 & -2 & -2 & -2 & -2 & -2 & 2 & 2 & -2 & -2 & 0 & 0 & 0 & 0 & 0 & 0 & 0 & 0 & 0 & 0 & 0 & 0 \\
$\chi_{20}$& 6 & -6 & 6 & 2 & -2 & 2 & -2 & 2 & -2 & 2 & -2 & 2 & 0 & 0 & 0 & 0 & 0 & 0 & 0 & 0 & 0 & 0 & 0 & 0 \\
$\chi_{21}$& 12 & 0 & -4 & -4 & 0 & 4 & 0 & 0 & 0 & 0 & 0 & 0 & 0 & 0 & -2 & -2 & 2 & 2 & 0 & 0 & 0 & 0 & 0 & 0 \\
$\chi_{22}$& 12 & 0 & -4 & -4 & 0 & 4 & 0 & 0 & 0 & 0 & 0 & 0 & 0 & 0 & 2 & 2 & -2 & -2 & 0 & 0 & 0 & 0 & 0 & 0 \\
$\chi_{23}$& 12 & 0 & -4 & 4 & 0 & -4 & 0 & 0 & 0 & 0 & 0 & 0 & 0 & 0 & -2 & 2 & -2 & 2 & 0 & 0 & 0 & 0 & 0 & 0 \\
 $\chi_{24}$&12 & 0 & -4 & 4 & 0 & -4 & 0 & 0 & 0 & 0 & 0 & 0 & 0 & 0 & 2 & -2 & 2 & -2 & 0 & 0 & 0 & 0 & 0 & 0 \\
\midrule
$[h]$&$1A$&$2B$& $2A$ & $\widehat{4F}$& $4B$ & $4B$ & $2A$ &$4B$&$2A$&$4B$&$2A$&$4B$&$3A$&$6A$&$2B$&$2A$&$4B$&$4B$&$8A$&$8A$&$4B$&$4B$&$4B$&$4B$  \\
\bottomrule
\end{tabular}
\label{tbl:quarticW4}
\end{footnotesize}
\end{center}
\end{sidewaystable}
\restoregeometry

\begin{table}
\begin{center}
\caption{Character table of $M_{11}$, with $\f_g = \phi^{12A_2}_{h}$.
$\b_{11}:= (-1 + \sqrt{-11})/2$  }
%\smallskip
\begin{footnotesize}
\begin{tabular}{c|rrrrrrrrrrr}\toprule
$[g]$&1A&2A&3A&4A&5A&6A&8A&8B&11A&11B\\
$[g^2]$&1A&1A&3A&2A&5A&3A&4A&4A&11B&11A\\
$[g^3]$&1A&2A&1A&4A&5A&2A&8A&8B&11A&11B\\
$[g^5]$&1A&2A&3A&4A&1A&6A&8B&8A&11A&11B\\
\midrule
${\chi}_{1}$&1&1&1&1&1&1&1&1&1&1\\
${\chi}_{2}$&10&2&1&2&0&-1&0&0&-1&-1\\
${\chi}_{3}$&10&-2&1&0&0&1&$\sqrt{2}i$&-$\sqrt{2}i$&-1&-1\\
${\chi}_{4}$&10&-2&1&0&0&1&-$\sqrt{2}i$&$\sqrt{2}i$&-1&-1\\
${\chi}_{5}$&11&3&2&-1&1&0&-1&-1&0&0\\
${\chi}_{6}$&16&0&-2&0&1&0&0&0&$\beta_{11}$&$\overline{\beta_{11}}$\\
${\chi}_{7}$&16&0&-2&0&1&0&0&0&$\overline{\beta_{11}}$&$\beta_{11}$\\
${\chi}_{8}$&44&4&-1&0&-1&1&0&0&0&0\\
${\chi}_{9}$&45&-3&0&1&0&0&-1&-1&1&1\\
${\chi}_{10}$&55&-1&1&-1&0&-1&1&1&0&0\\
\midrule
$\Pi_g$& $1^{12}$&  $1^42^4$&  $1^33^3$& $2^24^2$&$1^2 5^2$ &$1^12^13^16^1$&$4^18^1$&$4^18^1$&$1^1 11^1$&$1^1 11^1$ \\\midrule
$[h]$ & $1A$ & $2B$ &$3A$ & $4C$&$5A$ &  $6C$ &$8CD$&$8CD$ &$11AB$&$11AB$\\
\bottomrule
\end{tabular}
\end{footnotesize}
\label{tbl:M11}
\end{center}
\end{table}

\clearpage
\section{Twining Genera from the Conway Module}\label{sec:conwaytwinings}

The module analyzed in \cite{DuncanMackCrane} of the 4-plane preserving Conway subgroups  leads to proposed $K3$ twining genera given by
\begin{multline} 
\phi_g(\tau, z) = -{1 \over 2}\left({\th_4(\t, z)^2 \over \th_4(\t, 0)^2}{\eta_g(\tau/2) \over \eta_g(\tau)} - {\th_3(\t, z)^2 \over \th_3(\t, 0)^2}{\eta_{-g}(\tau/2) \over \eta_{-g}(\tau)} \right) \\ + {1 \over 2}\left({\th_1(\t, z)^2 \over \eta(\t)^6}D_g \eta_g(\t) - {\th_2(\t, z)^2 \over \th_2(\t, 0)^2}C_{-g}\eta_{-g}(\tau) \right) 
\end{multline}
where, for an element $g \in {\rm Co}_0$ with Frame shape $\pi_g = \prod_{m>0}(m)^{k_m}$, we define the eta-product
\be 
\eta_g(\tau) = \prod_{m>0}\eta(m \t)^{k_m}.
\ee
and $C_g, D_g$ are certain $g$-dependent constants. See \cite{DuncanMackCrane} for more details.

Note that when $\langle g \rangle$ leaves precisely a four-dimensional subspace invariant in the 24-dimensional representation of Conway, namely when there are precisely four of the twenty-four eigenvalues coinciding with 1, one has a choice of sign of $D_g$ which leads to different twining genera. As a result, to each of these conjugacy classes one can attach two  functions \cite{DuncanMackCrane}. In what follows and in the tables in the main text, we use the notation $\phi_{\widehat{g}}$, with $g$ given by the standard ${\rm Co}_0$ conjugacy names, to denote these functions. In the case where two different twining genera are attached to a given conjugacy classes, we use the notation $\phi_{\widehat{g}}$ and $\phi_{\widehat{g}'}$.

For completeness, in what follows we also give the 
 first few coefficients for the Conway twinings that appear in our models: 
\begin{align}\nonumber
\phi_{\widehat{3C}}(\t, z)&= ({2 \over y} - 7 + 2 y) + q\left({-7 \over y^2} + {61 \over y} - 108 + 61 y - 7 y^2\right) + O(q^2)\\\nonumber
\phi_{\widehat{4F}}(\t, z)&= ({2 \over y} - 8 + 2 y) + q\left({-8 \over y^2} + {64 \over y} - 112 + 64 y - 8 y^2\right) + O(q^2)\\\nonumber
\phi_{\widehat{6I}}(\t, z)&= ({2 \over y} + 1  + 2 y) + q\left({1 \over y^2} + {3 \over y} - 8 + 3 y +1  y^2\right) + O(q^2)\\\nonumber
\phi_{\widehat{9C}}(\t, z)&= ({2 \over y} - 1  + 2 y) + q\left({-1 \over y^2} + {1 \over y} +  y -1  y^2\right) + O(q^2)\\\nonumber
\phi_{\widehat{9C'}}(\t, z)&=({2 \over y} - 1  + 2 y) + q\left({-1 \over y^2} + {10 \over y} - 18 + 10 y  -1  y^2\right) + O(q^2) \\\nonumber
\phi_{\widehat{12L}}(\t, z)&= ({2 \over y} - 3  + 2 y)  + q\left({-3 \over y^2} + {12 - 3\sqrt{3} \over y} - 18 + 6 \sqrt{3} + (12 - 3\sqrt{3})y - 3  y^2\right) \\\nonumber & + O(q^2)\\\nonumber
\phi_{\widehat{12L'}}(\t, z)&= ({2 \over y} - 3  + 2 y)  + q\left({-3 \over y^2} + {12 + 3\sqrt{3} \over y} - 18 - 6 \sqrt{3} + (12 + 3\sqrt{3})y - 3  y^2\right) \\ \nonumber& + O(q^2).\\\nonumber
\end{align}

%\addcontentsline{toc}{section}{References}
%\bibliographystyle{alpha}
%\bibliographystyle{../../bib/utphys}
%\bibliography{../../bib/main_bib}

    \end{document}